%% file: DFT_NF_BA.tex
\documentclass[journal]{IEEEtran} % double column

\usepackage[utf8]{inputenc}
\usepackage{tabularx}
\usepackage{algorithm}
\usepackage{algorithmic}
\usepackage{array}
\usepackage{amsfonts}
\usepackage{multirow}
\usepackage{amssymb}
\usepackage{amsmath, amsthm}
\usepackage{mathtools}
\usepackage{bbm}
\usepackage{graphicx}
\usepackage{epstopdf}

\newcommand{\argmax}{\operatornamewithlimits{argmax}}
\usepackage{epsf}
\usepackage{amsmath, amsfonts}
\usepackage{latexsym}
\usepackage{subfigure}
\usepackage{multirow}
\usepackage{amssymb}
\usepackage{float}
\usepackage{enumerate}
\usepackage{cite}
\usepackage{color}
\usepackage{scrextend}
\usepackage{bbm}
\usepackage{lipsum}
\usepackage{mathtools}
\usepackage{cuted}
\usepackage{scalerel}
\usepackage{enumitem}
\allowdisplaybreaks

\begin{document}
%\bstctlcite{IEEEexample:BSTcontrol}

\title{DFT-based Near-field Beam Alignment: Model-based and Data-Driven Hybrid Approach}

\author{
    Hongjun~Heo,~\IEEEmembership{Student Member,~IEEE},
    and~Wan~Choi,~\IEEEmembership{Fellow,~IEEE}% <-this % stops a space
%    \thanks{XXX\emph{(Corresponding author: Wan Choi)}}
%    \thanks{H. Heo  and W. Choi are with the Department of Electrical and Computer Engineering, Seoul National University (SNU), Seoul 08826, Korea (e-mail:hhj2681@snu.ac.kr).}
    \thanks{ H. Heo and W. Choi are with the Department of Electrical and Computer Engineering, and the Institute of New Media and Communications, Seoul National University (SNU), Seoul 08826, Korea (e-mail: \{hhj2681, wanchoi\}@snu.ac.kr) \emph{(Corresponding author: Wan Choi).}}
}

\maketitle

% \vspace{-20pt}

\begin{abstract}    
Accurate beam alignment is a critical challenge in XL-MIMO systems, especially in the near-field regime, where conventional far-field assumptions no longer hold. Although 2D grid-based codebooks in the polar domain are widely accepted for capturing near-field effects, they often suffer from high complexity and inefficiency in both time and computational resources. To address this issue, we propose a novel line-of-sight (LoS) near-field beam alignment scheme that leverages the discrete Fourier transform (DFT) matrix, which is commonly used in far-field environments. This approach ensures backward compatibility with the legacy DFT codebook for far-field signals by allowing its reuse. By introducing a new method to analyze the energy spread effect, we define the concept of an $\epsilon$-approximated signal subspace, spanned by DFT vectors that exhibit significant correlation with the near-field channel vector. Building on this analysis, the proposed hybrid scheme integrates model-based principles with data-driven techniques. Specifically, it utilizes the properties of the DFT matrix for efficient coarse alignment while employing a deep neural network (DNN)-aided fine alignment process. The fine alignment operates within the reduced search space defined by the coarse alignment stage, significantly enhancing accuracy while reducing complexity. Simulation results demonstrate that the proposed scheme achieves superior alignment performance while reducing both computational and model complexity compared to existing methods.

\end{abstract}

\begin{IEEEkeywords}
Near-field, beam alignment, deep learning, XL-MIMO

\end{IEEEkeywords}

%\vspace{-0.5cm}
% \fi
\IEEEpeerreviewmaketitle
\newtheorem{mylemma}{Lemma}
\newtheorem{myremark}{Remark}
\newtheorem{mytheorem}{Theorem}
\newtheorem{mydef}{Definition}
\newtheorem{mycor}{Corollary}
\newtheorem{myexample}{Example}
\newtheorem{mydefinition}{Definition}
\newtheorem{myproposition}{Proposition}
\newcounter{mytempeqncnt}
\vspace{-0.5cm}
\section{Introduction}\label{sec:intro}
Extremely large-scale multiple-input multiple-output (XL-MIMO) systems are emerging as a key enabling technology for sixth-generation (6G) mobile communications. Expanding on the frequency ranges used in fifth-generation (5G) systems, XL-MIMO is expected to operate at even higher frequencies, including millimeter-wave and sub-terahertz bands, to meet the increasing demand for greater bandwidth~\cite{andrews20246G}. By leveraging a greater number of antennas and wider bandwidth, XL-MIMO achieves substantial increases in data rates, playing a critical role in realizing 6G's advanced objectives. However, the adoption of XL-MIMO brings about significant changes to wireless communication environments, necessitating adaptations to existing paradigms to effectively address these shifts.

One of the most significant changes is the near-field effect, caused by the enlarged array aperture and shorter wavelength at higher frequency bands. This effect necessitates a more accurate model of electromagnetic wave propagation, shifting from the conventional planar wave model, which assumes negligible signal variation across the aperture, to a spherical wave model that accounts for curvature in the wavefronts. Notably, spherical wavefront modeling in the near-field introduces additional resolvability in the distance or depth domain, enhancing energy focusing capabilities~\cite{bjornson2024enabling}.

To fully harness the benefits of XL-MIMO systems, selecting an appropriate beam prior to data transmission is essential~\cite{kang2024pilot, kim2025lowcomplexity}. Beam alignment or beam training, commonly used in 5G~\cite{andrews20145G}, relies on predefined codebooks to identify dominant signal paths and optimize high-gain beams. However, beam alignment schemes designed for massive MIMO in 5G primarily assume far-field channel models. When directly applied to near-field environments, these schemes suffer from significant performance degradation. Even in line-of-sight (LoS)-dominant environments—prevalent in high-frequency scenarios~\cite{yun2024analogdigital, palaiologos2024losnonuniform, zhang2022focus}—near-field channels lack the angular-domain sparsity typically captured by conventional angular-domain codebooks, such as the discrete Fourier transform (DFT) codebook, in far-field conditions. Specifically, while far-field channel vectors exhibit strong energy concentration at the corresponding angle when using a DFT codebook, near-field channel vectors display energy spread around the true angle. This lack of sparsity, caused by the \textit{energy spread} or \textit{dispersion effect}, introduces angular ambiguity when conventional codebooks like the DFT matrix are used~\cite{dai2022nfce}. As a result, energy detection with the DFT codebook becomes less effective.

To address these challenges,~\cite{dai2022nfce} proposed a polar-domain representation in which the near-field channel vector exhibits sparsity, along with a transform matrix that enables joint quantization of both angular and distance dimensions  along ellipses. Following this, numerous subsequent works introduced beam alignment schemes that use the polar-domain transform matrix as a codebook, testing each column as a potential codeword. For example,~\cite{cui2023rainbow} proposed an exhaustive search over the polar-domain codebook, using additional hardware to reduce the search time caused by the codebook's substantial size. However, this approach is unsuitable for legacy systems based on phased-array architectures. Alternatively,~\cite{qi2023dnbt} introduced a beam alignment scheme that tests only evenly sampled codewords from the polar-domain codebook and employs a deep neural network (DNN) to interpolate the pattern and map the relationships. However, the output dimension of the DNN matches the size of the polar-domain codebook, leading to a large computational footprint.  Additionally,~\cite{wu2024hier} adapted the concept of multi-layer hierarchical beam alignment, commonly used in far-field scenarios to progressively narrow beam width, for near-field applications with a polar-domain codebook to reduce search time. While effective in far-field environments, this method encounters challenges in high-frequency bands envisioned for 6G, where initial errors may arise because wide beams in higher layers cannot adequately compensate for significant path loss. Furthermore, unlike the resolution of spatial angles, the range resolution of an array is not proportional to the antenna aperture~\cite{ding2024resolution}, making hierarchical search an ineffective option when using a polar-domain codebook.

As explained above, most polar-domain codebook-based approaches focus on exploiting its sparsity characteristics while targeting a reduction in search time to manage the inherent high complexity of the codebook. However, these approaches often compromise other resources or fail to account for the unique characteristics of near-field channels. To reduce search time and complexity over polar-domain, some studies  sought to utilize DFT matrix as codebook used in far-field to identify dominant path by matching the energy spread pattern of the near-field signal~\cite{you2024dft}. This approach first determines the angle, based on the observation that the angle corresponds to the median of the energy spread, and subsequently estimates the range by numerically evaluating the width of the energy spread pattern. However, such pattern is highly non-linear, requiring complex numerical computations involving intricate integration and its inversion (or demapping) operations, which are numerical unstable and intractable. Furthermore, the process of determining the angle within the energy spread is based on an observation that lacks mathematical justification. Alternatively, DNNs was employed to model these non-linear relationships~\cite{wang2023naivecnn}. While DNNs demonstrate potential for accurate modeling, the approaches based on end-to-end neural network typically require large-scale networks, resulting in considerable computational burdens.

In this paper, we explore a novel near-field beam alignment scheme for LoS-dominant scenarios that utilizes the DFT codebook. As previously discussed, near-field beam alignment based on the DFT codebook can circumvent the high complexity associated with polar-domain codebook-based approaches. Additionally, it ensures backward compatibility with the legacy DFT codebook for far-field signals, allowing for its reuse. However, it suffers from angular ambiguity due to energy spread. To address the challenges introduced by the energy spread effect, we propose a structured and well-founded model that mitigates the resulting angular ambiguity. Our approach is based on analyzing the structural properties of the near-field channel vector in the DFT angular domain, particularly its energy spread. Specifically, we focus on the approximated signal subspace spanned by DFT vectors that exhibit strong correlation with the near-field channel vector. The proposed method begins with a coarse estimation of the signal subspace using a mathematical approach. This estimation is then refined through a neural network-based fine-tuning process, which operates within a reduced search space identified by the initial coarse estimation. Unlike traditional numerical methods or purely end-to-end neural network models, our hybrid approach integrates model-based techniques with data-driven strategies. This integration significantly reduces model complexity while maintaining high efficiency and accuracy in beam alignment.

The rest of the paper is organized as follows. Section~\ref{sec:systemmodel} presents the system and channel model. In Section~\ref{subsec:3-a}, we present a detailed mathematical analysis of the energy spread effect. Then, Section~\ref{subsec:3-b} defines the $\epsilon$-approximated signal subspace, which provides a quantitative model for energy spread. Based on this model, we propose a hybrid near-field beam alignment scheme in Section~\ref{subsec:3-c}, incorporating neural networks to improve performance. To evaluate the effectiveness of the proposed approach, extensive simulations are conducted, examining both beam alignment performance in  Section~\ref{subsec:BA_performance} and computational complexity in Section~\ref{subsec:complexity}. Finally, conclusions are drawn in Section~\ref{sec:conclusion}.

\textbf{\emph{Notations}}: The following notations are used throughout the paper. The imaginary unit is denoted by $j$ ($j = \sqrt{-1}$). The sets of real and complex numbers are $\mathbb{R}$ and $\mathbb{C}$, respectively. For a complex scalar, $|\cdot|$ denotes magnitude, and $\|\cdot\|$ represents the Euclidean (L2) norm of a vector. The real part of a scalar or vector is $\mathrm{Re}(\cdot)$. The superscript $^T$ indicates the transpose, and $^H$ represents the complex conjugate transpose. The identity matrix of size $p \times p$ is written as $\mathbf{I}_{p \times p}$. Gaussian random vectors are denoted as $\mathcal{N}(\mathbf{\mu}, \mathbf{\Sigma})$ for real vectors and $\mathcal{CN}(\mathbf{\mu}, \mathbf{\Sigma})$ for complex vectors, where $\mathbf{\mu}$ and $\mathbf{\Sigma}$ are the mean and covariance matrix, respectively. Bold lowercase font (e.g., $\mathbf{v}$) denotes vectors, bold uppercase font (e.g., $\mathbf{M}$) denotes matrices, and bold uppercase font with an arrow (e.g., $\vec{\mathbf{P}}$) specifies position vectors in real space. The notation $\{ f(k) \}_{k=a}^{b}$ represents a sequence indexed by integers $k = a, a+1, \ldots, b$. The floor function $\lfloor \cdot \rfloor$ and modulo operation $(a \:\mathrm{mod}\: b)$ are used to indicate the largest integer less than or equal to a given value and the remainder when dividing $a$ by $b$, respectively.

\section{System Model}\label{sec:systemmodel}
\subsection{Channel Model} \label{subsec:channel}
\begin{figure}[!t]
    \centering
    \includegraphics[width = 0.9\columnwidth]{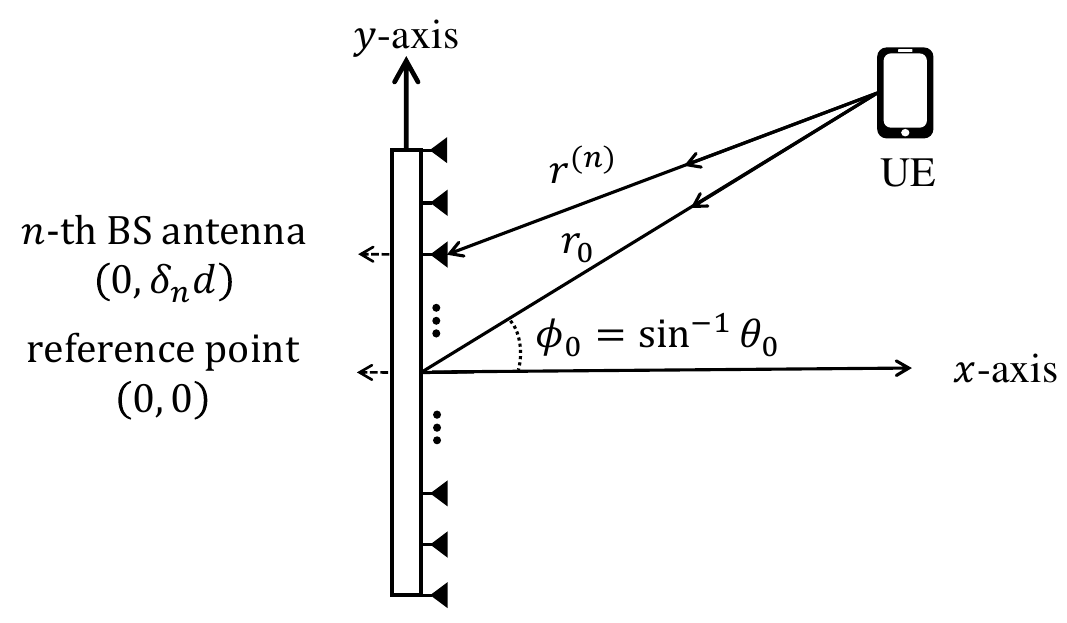}
  %  \vspace{-10pt}
    \caption{Illustration of system model and antenna configurations.}
    \label{fig:system_model}
    %\vspace{-10pt}
\end{figure}

Consider an uplink beam alignment for the XL-MIMO communication system where the single-antenna UE transmits known pilot symbols to the BS equipped with a uniform linear array (ULA) consisting of $N$ antennas connected to a single RF-chain. The BS performs analog beamforming, with the constraint that only one measurement can be taken at each measurement time. While the analysis focuses on analog beamforming, the proposed scheme can also be extended to hybrid digital-analog structures. We assume that the UE is located within the radiative near-field region, specifically at a distance greater than the Fresnel distance $d_F = 0.62\sqrt{D^3/\lambda}$ but less than the Rayleigh distance $d_R = 2D^2/\lambda$~\cite{polk1956fresnel}. Here, $D$ and $\lambda$ denote the array aperture size and carrier wavelength, respectively. With a half-wavelength antenna spacing $d=\lambda/2$, array aperture size satisfies $D=(N-1)d=\frac{(N-1)\lambda}{2}$. The $n$-th antenna at BS is located at $(0, \delta_n d)$ along the y-axis where $\delta_n = \frac{2n-N-1}{2}, n=1,...,N$, as shown in Fig.~\ref{fig:system_model}.

We consider the LoS channel from the UE to the BS, a common scenario in high-frequency bands~\cite{yun2024analogdigital, palaiologos2024losnonuniform, zhang2022focus}. With the aforementioned system, the LoS channel vector $\mathbf{h} \in \mathbb{C}^{N \times 1}$ is given by
\begin{align}
    \mathbf{h} = \sqrt{N} h_0 e^{-j \frac{2 \pi}{\lambda} r_0} \mathbf{a}(\theta_0, r_0), \label{eqn:channel}
\end{align}
where $h_0 = \frac{\lambda}{4 \pi r_0}$ is the path loss, $r_0$ is the distance between the reference point $(0, 0)$ of the BS and the UE, and $\theta_0 = \sin{\phi_0}$ denotes the spatial angle of the UE. Here, $\phi_0 \in \left[ -\pi/2, \pi/2 \right]$ is the physical angle in radian, and therefore, $\theta_0 \in \left[ -1, 1 \right]$. The near-field steering vector $\mathbf{a} \in \mathbb{C}^{N \times 1}$ captures the phase shifts across the array due to the near-field effects and is defined as
\begin{align}
	\mathbf{a}(\theta_0, r_0) = \frac{1}{\sqrt{N}} \left [ e^{-j \frac{2 \pi}{\lambda} (r_0^{(1)}-r_0)},...,e^{-j \frac{2 \pi}{\lambda} (r_0^{(N)}-r_0)}  \right ]^H, \label{eqn:nf_manifold}
\end{align}
where $r_0^{(n)}=\sqrt{r_0^2+\delta_n^2d^2-2 r_0 \delta_n d \theta_0}$ denotes the distance between the $n$-th BS antenna element and the UE. 

\subsection{DFT Combiner and Signal Model} \label{subsec:signal}
During the beam alignment procedure, the BS employs a predefined $N$-point DFT matrix, $\mathbf{F} = \left[\mathbf{f}(\theta_1), \dots, \mathbf{f}(\theta_N)\right] \in \mathbb{C}^{N \times N}$, as its codebook for combining. At the $m$-th measurement time slot, the $m$-th column of $\mathbf{F}$, denoted by $\mathbf{f}(\theta_m) \in \mathbb{C}^{N \times 1}$, is used as the combiner and is expressed as
\begin{align}
    \mathbf{f}(\theta_m) = \frac{1}{\sqrt{N}} \left[e^{j \frac{2 \pi}{\lambda} \delta_1 d \theta_m}, \dots, e^{j \frac{2 \pi}{\lambda} \delta_N d \theta_m}\right]^H, \label{eqn:DFT}
\end{align}
where $\theta_m = \frac{2m - N - 1}{N}$ $\forall m = 1, \dots, N$ represents the $m$-th sample of the spatial angle, uniformly spaced over the angular-domain $\left[ -1, 1 \right]$. With the definitions of $\delta_n$ and $\theta_m$, and after some calculations, it can be verified that $\omega^{\left( n-\frac{N+1}{2} \right) \left( m-\frac{N+1}{2} \right)}$ represents the DFT kernel where $\omega = e^{-j \frac{2 \pi}{N}}$ is a primitive $N$-th root of unity.

The received signal at the BS, combined using the beam $\mathbf{f}(\theta_m)$ at the $m$-th measurement time slot, can be expressed as
\begin{align}
    y_m = \sqrt{P_t} \mathbf{f}(\theta_m)^H \mathbf{h}s + \mathbf{f}(\theta_m)^H \mathbf{w}_m, \label{eqn:rx_signal}
\end{align}
where $P_t$ denotes the power of the uplink transmit power of the UE, $s$ is unit-power pilot symbol, i.e., $\left| s \right|^2 = 1$. For simplicity, we assume $s = 1$. The noise vector $\mathbf{w}_m \sim \mathcal{CN}(0, \sigma^2 \mathbf{I}_{N \times N})$ represents independent Gaussian noise. When the received signal is combined using the DFT beam $\mathbf{f}(\theta_m)$, the resulting effective noise term after combining, $z_m = \mathbf{f}(\theta_m)^H \mathbf{w}_m$, is reduced to a scalar additive white Gaussian noise (AWGN) with distribution $\mathcal{CN}(0, \sigma^2)$.

By stacking all $N$ measurements, the received signal can be represented in matrix form as
\begin{align}
    \mathbf{y} = \sqrt{P_t} \mathbf{F}^H \mathbf{h} + \mathbf{z}, \label{eqn:stacked_rx_signal}
\end{align}
where $\mathbf{y} \in \mathbb{C}^{N \times 1}$ is the stacked received signal vector, and $\mathbf{z} \in \mathbb{C}^{N \times 1}$ is the effective noise vector.

It is well-known that the DFT matrix for beam alignment offers several advantages under the assumption of far-field channel modeling. First, by definition, combining with the DFT inherently transforms the spatial domain into the angular-domain, where the high frequency far-field channel model exhibits sparsity. Second, each column of the DFT matrix corresponds to a far-field steering beam with a phase that varies linearly with the antenna indices, enabling efficient scanning of the entire angular-domain using predefined orthogonal beams. These properties simplify the beam alignment procedure into a straightforward energy detection problem in the angular-domain.

However, in the near-field scenario, the channel vector does not exhibit sparsity in the angular-domain, leading to \textit{energy spread} or \textit{dispersion effect}~\cite{dai2022nfce}, which causes ambiguity in angle estimation. Additionally, the need to account for range significantly increases the complexity of the problem. To address this, recent studies have explored the use of polar domain codebook~\cite{dai2022nfce}, which transform the near-field channel vector into a sparse representation in the polar-domain. In the polar-domain codebook, there are $N$ uniformly sampled angles, each paired with non-uniformly sampled $Q$ discrete range samples. The parameter $Q$ is a design variable that controls the column coherence of the transform matrix
\begin{equation}
    \begin{aligned}
        \mathbf{F}_{\text{Polar}} = \big[ & \mathbf{a}(\theta_1, r_{1,1}), \dots, \mathbf{a}(\theta_1, r_{1,Q}), \dots, \\
        & \mathbf{a}(\theta_N, r_{N,1}), \dots, \mathbf{a}(\theta_N, r_{N,Q}) \big] \in \mathbb{C}^{N \times NQ}, \label{eqn:polar_trans}
    \end{aligned}
\end{equation}
where $\mathbf{a}(\theta_m, r_{m,q})$ denotes the steering vector corresponding to the 
$m$-th angle and $q$-th range sample. This structure increases the dimension of the transform matrix and requires testing $Q$-times more codewords compared to the DFT matrix, making it computationally expensive. For instance, with $256$ antenna elements operating at $28$ GHz, achieving a column coherence below $0.7$ for the same angular samples necessitates setting $Q$ to approximately $16$.

Although the polar-domain codebook effectively transforms the near-field channel vector into a sparse representation, its increased complexity poses challenges for practical implementation.  To address this, leveraging the structural advantages of the DFT matrix, combined with near-field adaptations, offers a promising alternative. In the next section, we introduce a novel approach that harnesses the inherent properties of the DFT matrix while refining its structure to better accommodate near-field conditions.

\section{Proposed Scheme}\label{sec:proposed}
To utilize the DFT matrix for beam alignment in the near-field environment, we first analyze the near-field manifold in the angular-domain. Based on it, we preprocess the received signals as input for machine learning-based fine-tuning.

\subsection{Angular-Domain Representation} \label{subsec:3-a}
To describe the behavior of the energy spread effect, consider the normalized near-field channel vector $\tilde{\mathbf{h}} = \mathbf{a}(\theta_{\zeta_0}, r_0)$ from~\eqref{eqn:nf_manifold}, where $\theta_{\zeta_0}$ is the spatial angle, and 
\begin{align}
\zeta_0 = \underset{m \in \{1, \dots, N\}}{\arg\min} |\theta_0 - \theta_m|. 
\end{align}
That is, $\theta_{\zeta_0}$ is the closest spatial sample of $\theta_0$ in the DFT codebook grid and $\zeta_0$ is its index. The squared magnitude of the correlation between $\tilde{\mathbf{h}}$ and the DFT beam $\mathbf{f}(\theta_{\zeta_0, l})$, whose phase is steered toward the direction $\theta_{\zeta_0, l} = \theta_{\zeta_0} + \frac{2l}{N}$ for an integer $l$, is expressed as
\begin{align}
    \rho& \left( \theta_{\zeta_0}, r_0 ; \frac{2l}{N} \right) 
    = \left| \mathbf{f} (\theta_{\zeta_0, l})^H \tilde{\mathbf{h}} \right|^2 \notag \\
    =& \frac{1}{N^2} \left| \sum_{n=1}^{N} e^{j \frac{2 \pi}{\lambda} 
    \left( r_{\zeta_0}^{(n)} - r_0 + \delta_n d \theta_{\zeta_0, l} \right)} \right|^2 \notag \\
    \overset{(a)}{\approx}& \frac{1}{N^2} \left| \sum_{n=-\frac{N-1}{2}}^{\frac{N-1}{2}} e^{j \pi s \left( n + \frac{l}{N s} \right)^2} \right|^2 \notag\\
    \overset{(b)}{\approx}& \frac{1}{\Delta^2} \Bigg| 
    \int_{w}^{w+\Delta} e^{j x^2} \mathrm{d}x \Bigg|^2 \notag \\
    \overset{(c)}{=}& \frac{1}{\Delta^2} \Bigg[ \Big( C\big( w + \Delta \big) - C\big( w \big) \Big)^2 + \Big( S\big( w + \Delta \big) - S\big( w \big) \Big)^2 \Bigg] \label{eqn:correlation}
\end{align}
where (a) is derived using the Fresnel approximation $r_{\zeta_0}^{(n)} - r_0 \approx -\theta_{\zeta_0} \delta_n d + \frac{\delta_n^2 d^2 (1 - \theta_{\zeta_0}^2)}{2 r_0}$ due to the fact that communication distance $r_0$ is much larger than the size of aperture $D$, $D \ll r_0$ and a change of variable $s = \frac{d(1 - \theta_{\zeta_0}^2)}{2 r_0}$. In (b), the summation is approximated as a Fresnel integral with the change of variables $w = \sqrt{\pi s} \big( \frac{l}{N s} - \frac{N}{2} \big) \quad \text{and} \quad \Delta = N \sqrt{\pi s}.$ Finally, (c) employs the Fresnel cosine and sine integrals, defined as $C(x) = \int_{0}^{x} \cos(t^2) \, \mathrm{d}t \quad \text{and} \quad S(x) = \int_{0}^{x} \sin(t^2) \, \mathrm{d}t.$ Note that $\frac{2l}{N}$ corresponds to the spatial angle deviation from $\theta_{\zeta_0}$ in the angular-domain. The integer $l$ indicates the extent of this deviation in terms of the number of beams, with a beamwidth or spatial resolution of $\frac{2}{N}$.

While previous works such as~\cite{dai2022nfce, you2024dft, heath2023widenear} proposed similar forms of correlation functions, none have leveraged the discretized angular properties of DFT to describe the behavior of $\rho \left(\theta_{\zeta_0}, r_0 ; \frac{2\tilde{l}}{N}\right)$ in~\eqref{eqn:correlation}. Specifically, they either treated the argument $w$ of $\rho \left(\theta_{\zeta_0}, r_0 ; \frac{2\tilde{l}}{N}\right)$ as continuous or failed to fully exploit its discrete nature and associated properties. Furthermore, it is well-known that the Fresnel integrals in~\eqref{eqn:correlation} cannot be expressed in terms of elementary functions. As a result, the prior studies have  resorted to numerical search methods. However, as emphasized in~\cite{heath2023widenear},  a comprehensive understanding of the inverse mapping in~\eqref{eqn:correlation} and its efficient evaluation is crucial from a signal processing perspective. Numerical search methods, however, are inadequate for deriving this inverse mapping.

\subsection{Approximated Signal Subspace} \label{subsec:3-b}
Given the periodicity of the complex exponential with respect to the integer $l$ with period $N$ in~\eqref{eqn:correlation}, the set of functions $\left\{ \rho \left(\theta_{\zeta_0}, r_0 ; \frac{2l}{N}\right) \right\}_{l=-\frac{N}{2}+1}^{\frac{N}{2}}$ represents the energy of the angular-domain samples of the near-field channel vector. Due to the symmetry about $l=0$, which arises from the odd symmetry of the Fresnel integrals, $C(-x) = -C(x)$ and $S(-x) = -S(x)$, we focus on $l = 1, \dots, N/2$ for simplicity. Additionally, while $\theta_{\zeta_0, l} = \theta_{\zeta_0} + \frac{2\tilde{l}}{N}$ may fall outside the range of spatial angle $[-1, 1]$ for some $\zeta_0$ and $\tilde{l} \in \left\{ 1, \dots, N/2 \right\}$, the DFT beam index with the same correlation, $\rho \left(\theta_{\zeta_0}, r_0 ; \frac{2\tilde{l}}{N}\right) = \rho \left(\theta_{\zeta_0}, r_0 ; \frac{2I(\tilde{l})}{N}\right)$, can be found using the index mapping function $I(l) = \left((\zeta_0 + l - 1) \:\mathrm{mod} \:N\right) + 1$. Therefore, for notational simplicity, we assume throughout the paper that $\theta_{\zeta_0, l} = \theta_{\zeta_0} + \frac{2l}{N} \in [-1, 1]$ for $l \in \left\{ 1, \dots, N/2 \right\}$.

In the derivation, $\Delta^2 \rho \left(\theta_{\zeta_0}, r_0 ; \frac{2l}{N}\right)$ represents the squared Euclidean distance between $\left( C\left( w \right), S\left( w \right) \right)$ and $\left( C\left( w + \Delta \right), S\left( w + \Delta \right) \right)$ in $\mathbb{R}^2$ for $w > 0$. Note that the point $\left( C(t), S(t) \right)$, parameterized by $t$, forms a spiral known as the \textit{Cornu} or \textit{Euler spiral}. The following lemmas provide the useful properties of Cornu spiral to describe the behavior of $\rho \left(\theta_{\zeta_0}, r_0 ; \frac{2l}{N}\right)$.
\begin{figure}[!t]
    \centering
    \includegraphics[width = 1\columnwidth]{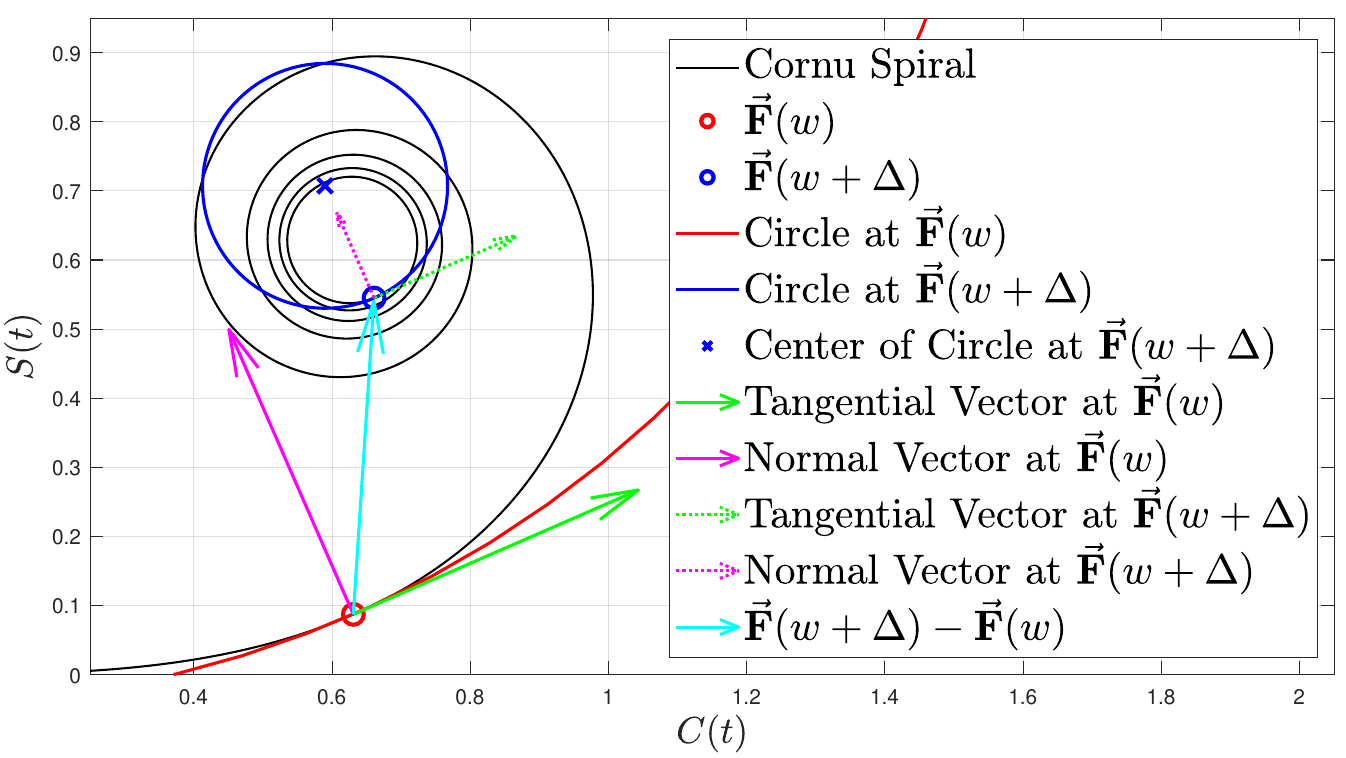}
    \caption{Two sample points, $ \vec{\mathbf{F}}(w) $ and $ \vec{\mathbf{F}}(w+\Delta) $, on the Cornu spiral with $ w = 0.6416 $ and $ \Delta = 5 $, along with their osculating circles. The center of the larger circle, which is not shown in the figure, is located at $ (0.0072, 1.5154) $.}
    \label{fig:spiral}
\end{figure}

\begin{mylemma}
    Let $\vec{\mathbf{F}}(t) := \big( C(t), S(t) \big)$ represent a point on the Cornu spiral, where $C(t)$ and $S(t)$ are the Fresnel integrals. The tangential and normal directions at $\vec{\mathbf{F}}(t)$ are given by $\vec{\mathbf{T}}(t) \!=\! \big( \cos(t^2), \sin(t^2) \big) \quad \text{and} \quad \vec{\mathbf{N}}(t) \!=\! \big( -\sin(t^2), \cos(t^2) \big),$
    respectively. Furthermore, the curvature and the radius of curvature at $\vec{\mathbf{Z}}(t)$ are $\kappa(t) \!=\! 2t \quad \text{and} \quad R(t) \!=\! \frac{1}{\kappa(t)} \!=\! \frac{1}{2t},$
    respectively.
    \label{lemma:properties}
\end{mylemma}

%\vspace{-5pt}

\begin{IEEEproof}
These are directly derived from the definition of Fresnel integrals and their first and second derivatives.
\end{IEEEproof}

\begin{mylemma}
    \textnormal{\cite{stoer1982curve}} Let $t > \bar{t} > 0$. Consider the Cornu spiral after $t$, defined as $\vec{\mathbf{Z}}(t') \!:=\! \left\{ \vec{\mathbf{F}}(t') \!=\! \big( C(t'), S(t') \big) \right\}$ for $\forall t' \!\geq\! t$, where $C(t')$ and $S(t')$ are the Fresnel integrals. Let the osculating circle at $\vec{\mathbf{F}}(t)$ be the circle centered at the center of curvature with a radius equal to the radius of curvature~\cite{stoker2011differential}. Then, the Cornu spiral $\vec{\mathbf{Z}}(t')$ for $\forall t' \geq t$ and the osculating circle at $\vec{\mathbf{F}}(t)$ are entirely contained within the interior of the osculating circle at $\vec{\mathbf{F}}(\bar{t})$.
    \label{lemma:osculating}
\end{mylemma}

\begin{IEEEproof}
See the reference~\cite{stoer1982curve}.
\end{IEEEproof}

The properties of \textbf{Lemmas~\ref{lemma:properties}} and \textbf{\ref{lemma:osculating}} are illustrated in detail in Fig.~\ref{fig:spiral}. Using \textbf{Lemmas~\ref{lemma:properties}}, \textbf{\ref{lemma:osculating}} and~\eqref{eqn:correlation}, we establish the following theorem regarding the upper bound of $\rho \left(\theta_{\zeta_0}, r_0 ; \frac{2l}{N}\right)$.

\begin{mytheorem}
    For $\Delta \!=\! N\sqrt{\pi s} \!>\! 0$ and $w \!=\! \sqrt{\pi s} \left( \frac{l}{N s} \!-\! \frac{N}{2} \right) \!>\! 0$, where $l$ is a positive integer and $s = \frac{d(1 - \theta_{\zeta_0}^2)}{2 r_0}$, the squared correlation $\rho \left( \theta_{\zeta_0}, r_0 ; \frac{2l}{N} \right)$ is upper bounded by
    $$
        \rho \left( \theta_{\zeta_0}, r_0 ; \frac{2l}{N} \right) < \frac{1}{\Delta^2} \left( \frac{1}{w} - \frac{1}{w + \Delta} \right)^2=\frac{1}{w^2(w+\Delta)^2}.
    $$
    \label{theorem:bound}
\end{mytheorem}

\begin{IEEEproof}
Consider two points $\vec{\mathbf{F}}(w) = \left( C(w), S(w) \right)$ and $\vec{\mathbf{F}}(w+\Delta) = \left( C(w+\Delta), S(w+\Delta) \right)$ on the Cornu spiral defined as $\vec{\mathbf{Z}}(t) \!:=\! \left\{ \vec{\mathbf{F}}(t) \!=\! \big( C(t), S(t) \big) \right\}$ for $\forall t \!\geq\! 0$. The centers of the osculating circles at $\vec{\mathbf{F}}(w)$ and $\vec{\mathbf{F}}(w+\Delta)$ are given by $\vec{\mathbf{M}}(w) = \vec{\mathbf{F}}(w) + R(w)\vec{\mathbf{N}}(w)$ and $\vec{\mathbf{M}}(w+\Delta) = \vec{\mathbf{F}}(w+\Delta) + R(w+\Delta)\vec{\mathbf{N}}(w+\Delta)$, respectively. The Euclidean distance between $\vec{\mathbf{F}}(w)$ and $\vec{\mathbf{F}}(w+\Delta)$ is obtained as
\begin{align}
    \Big\| &\vec{\mathbf{F}}(w+\Delta) - \vec{\mathbf{F}}(w) \Big\| 
    = \Big\| \Big( \vec{\mathbf{M}}(w+\Delta) - \vec{\mathbf{M}}(w) \Big) \notag \\
    &\hspace{6.55em} + \Big(   
    R(w)\vec{\mathbf{N}}(w) \!-\! R(w+\Delta)\vec{\mathbf{N}}(w+\Delta) \Big) \Big\| \notag\\
    &\overset{(a)}{=}\! \Big\| \Big( \vec{\mathbf{M}}(w+\Delta) \!-\! \vec{\mathbf{M}}(w) \Big) \!+\! \Big( R(w) \!-\! R(w+\Delta) \Big) \vec{\mathbf{N}}(w) \Big\| \notag \\
    &\overset{(b)}{\leq} \Big\| \vec{\mathbf{M}}(w+\Delta) \!-\! \vec{\mathbf{M}}(w) \Big\| 
    \!+\! \Big\| \Big( R(w) \!-\! R(w+\Delta) \Big) \vec{\mathbf{N}}(w) \Big\| \notag\\
    &\overset{(c)}{\leq} 2\Big| R(w) \!-\! R(w+\Delta) \Big| 
    = \left( \frac{1}{w} - \frac{1}{w + \Delta} \right)
\end{align}
where (a) follows from the fact that $\vec{\mathbf{N}}(w) = \vec{\mathbf{N}}(w+\Delta)$, as $w^2 = \frac{\pi^2 l^2}{\Delta^2} + \frac{\Delta^2}{4} - \pi l = \left( \frac{\pi^2 l^2}{\Delta^2} + \frac{\Delta^2}{4} + \pi l \right) - 2 \pi l = (w+\Delta)^2 - 2\pi l$, where $l$ is an integer, and \textbf{Lemma~\ref{lemma:properties}} is applied; (b) follows by applying triangle inequality; (c) is derived from \textbf{Lemma~\ref{lemma:osculating}}, which states that the osculating circle at $\vec{\mathbf{F}}(w+\Delta)$ is contained in the interior of the osculating circle at $\vec{\mathbf{F}}(w)$ and thus, the Euclidean distance between the two centers is smaller than the difference of the radii and $ \left\| \vec{\mathbf{N}}(w) \right\| =1$; and the last equality is from the definition of the radius of curvature of Cornu spiral from \textbf{Lemma~\ref{lemma:properties}}.
Combining this inequality and~\eqref{eqn:correlation}, we have 
$2|R(w) - R(w+\Delta)| = \left( \frac{1}{w} - \frac{1}{w + \Delta} \right) \geq \Big\| \vec{\mathbf{F}}(w+\Delta) - \vec{\mathbf{F}}(w) \Big\| = \Delta^2 \rho \left(\theta_{\zeta_0}, r_0 ; \frac{2l}{N}\right)$ and establish \textbf{Theorem~\ref{theorem:bound}}.
\end{IEEEproof}

From the perspective of the DFT beam, we establish the following corollary,

\begin{mycor} 
    For $\Delta \!=\! N\sqrt{\pi s} \!>\! 0$ and $w \!=\! \sqrt{\pi s} \left( \frac{l}{N s} \!-\! \frac{N}{2} \right) \!>\! 0$, with a small positive $\epsilon$ satisfying $$ 0 < \frac{1}{\pi s (1 - N^2 s^2)} = \frac{1}{w(w+\Delta)} \bigg|_{l=N/2} < \epsilon \ll 1,$$ where $l$ is a positive integer and $s = \frac{d(1 - \theta_{\zeta_0}^2)}{2 r_0}$,\\
    if $\frac{N}{2} > l' > \sqrt{\frac{\Delta^2}{2\pi} \left( \frac{\Delta^2}{2\pi} + \frac{2}{\pi\epsilon} \right)}$, then $$\rho \left(\theta_{\zeta_0}, r_0; \frac{2l'}{N}\right) < \epsilon^2.$$
    \label{cor:bound2}
\end{mycor}

\begin{IEEEproof}
From \textbf{Theorem~\ref{theorem:bound}}, we upper bound $\frac{1}{\Delta^2} \left( \frac{1}{w} \!-\! \frac{1}{w\!+\!\Delta} \right)^2 \!=\! \frac{1}{w^2(w\!+\!\Delta)^2}$ by $\epsilon^2$, yielding the desired result.
\end{IEEEproof}

Note that $0 < \frac{1}{\pi s (1 - N^2 s^2)}$ is always satisfied, since we have already assumed $D \ll r_0$, and therefore $\frac{D(1-\theta_{\zeta_0}^2)}{2} < D < r_0$, which implies $N s < 1$.

\textbf{Corollary~\ref{cor:bound2}} implies that if the angular deviation of the DFT beam from the true UE's angle is at least 
$\frac{2}{N}\sqrt{\frac{\Delta^2}{2\pi} \left( \frac{\Delta^2}{2\pi} + \frac{2}{\pi\epsilon} \right)}$, 
then the correlation is less than $\epsilon$. In other words, the energy of the corresponding angular-domain sample is less than $\epsilon^2$.

Using above results and the symmetry of $\rho$ with respect to $l=0$, we define the $\epsilon$-approximated signal subspace.

\begin{mydef}
Consider a UE located at a spatial angle~$\theta$~and a distance~$r$. Given the parameters $s = \frac{d(1 - \theta^2)}{2r}, \Delta = N \sqrt{\pi s}$ and $\epsilon$ satisfying $\frac{1}{\pi s (1 - N^2 s^2)} < \epsilon \ll 1$, 
the \textbf{$\epsilon$-approximated signal subspace}, denoted as~$\mathbf{V}_\epsilon$, is spanned by the following orthonormal basis vectors:
\begin{align}
    \begin{aligned}
        \Bigg[&
        \mathbf{f}\left( \theta - \frac{2}{N}L(\Delta, \epsilon) \right),
        \mathbf{f}\left( \theta - \frac{2}{N}(L(\Delta, \epsilon) - 1) \right), \dots, \\
        &\mathbf{f}\left( \theta + \frac{2}{N}(L(\Delta, \epsilon) - 1) \right),
        \mathbf{f}\left( \theta + \frac{2}{N}L(\Delta, \epsilon) \right)
        \Bigg],\label{eqn:signalspace}
    \end{aligned}
\end{align}
where 
\begin{align}
    L(\Delta, \epsilon) = \left\lfloor \sqrt{\frac{\Delta^2}{2\pi} \left( \frac{\Delta^2}{2\pi} + \frac{2}{\pi\epsilon} \right)} \right\rfloor,
\end{align}
and $\mathbf{f}(\cdot)$ represents the columns of the DFT matrix, as specified in~\eqref{eqn:DFT}.
\label{def:signalspace}
\end{mydef}

It is important to note that projecting the near-field manifold onto any of the basis vectors of $\mathbf{V}_\epsilon$ in~\eqref{eqn:signalspace} does not always yield a correlation greater than $\epsilon$. However, projecting onto any basis vector of its orthogonal complement $\mathbf{V}_\epsilon^{\perp}$, i.e., the DFT columns other than those in~\eqref{eqn:signalspace}, ensures that the resulting correlation is strictly less than $\epsilon$. This implies that, for a sufficiently small $\epsilon \ll 1$ that satisfies the condition in \textbf{Corollary~\ref{cor:bound2}}, the majority of the energy in the near-field manifold is concentrated in the projections onto the DFT matrix columns associated with spatial angles close to the true spatial angle.

The preceding discussion provides a valuable mathematical characterization of the energy spread effect. In the next subsection, we introduce a novel beam alignment scheme that leverages this characterization.

\subsection{Two-stage Beam Alignment} \label{subsec:3-c}
In this subsection, we propose the two-stage beam alignment process based on the analyzed properties of near-field manifold in angular-domain. The process is divided into two-stages.

\subsubsection{Coarse Alignment}
The objective of this process is to identify the $\epsilon$-approximated signal subspace, $\mathbf{V}_{\epsilon}$, derived from the received signals combined with DFT beams. This involves pinpointing the measurements where significant energy is concentrated. However, determining $\mathbf{V}_{\epsilon}$---as defined in~\eqref{eqn:signalspace}---is nontrivial because it varies with both $\theta_0$ and $r_0$, making it difficult to simultaneously ascertain its cardinality and position. To address this challenge, we exploit the unitary property of the DFT matrix to estimate the communication distance $r_0$ from the signal energy. Specifically, from~\eqref{eqn:stacked_rx_signal}, we can deduce that
\begin{align}
    \left\| \mathbf{y} \right\|^2 
    &= \left\| \sqrt{P_t} \mathbf{F}^H \mathbf{h} + \mathbf{z} \right\|^2 \notag \\
    &\overset{(a)}{=} P_t \left\| \mathbf{h} \right\|^2 \!+\! \left\| \mathbf{z} \right\|^2 
    \!+\! \sqrt{P_t} \left( \mathbf{F}^H \mathbf{h} \right)^H \mathbf{z} 
    \!+\! \sqrt{P_t} \mathbf{z}^H \left( \mathbf{F}^H \mathbf{h} \right) \notag \\
    &\overset{(b)}{=} P_t \left\| \mathbf{h} \right\|^2 \!+\! \left\| \mathbf{z} \right\|^2 
    \!+\! 2\sqrt{P_t} \mathrm{Re}\left\{ \left( \mathbf{F}^H \mathbf{h} \right)^H \mathbf{z} \right\} \notag \\
    &\overset{(c)}{\approx} P_t \left\| \mathbf{h} \right\|^2 \!+\! N\sigma^2 
    \!+\! 2\sqrt{P_t} \mathrm{Re}\left\{ \left( \mathbf{F}^H \mathbf{h} \right)^H \mathbf{z} \right\} \notag \\
    &\overset{(d)}{=} P_t \left( \left\| \mathbf{h} \right\|^2 \!+\! \frac{N\sigma^2}{P_t} \!+\! \mathcal{N}\left(0, \frac{2 \left\| \mathbf{h} \right\|^2 \sigma^2}{P_t}\right) \right), \label{eqn:rx_power}
\end{align}
where (a) is because of Parseval's theorem;  (b) is obtained from the relation $q + q^H = 2\mathrm{Re}\{q\}$; (c) is due to the i.i.d. assumption of noise and the law of large numbers,  $\left\| \mathbf{z} \right\|^2/N = \sum_{m=1}^{N} |z_m|^2/N \to \mathbb{E}\left\{\left( z_m-\mathbb{E}\left\{ z_m \right\} \right)^2 \right\}$ as $N \to \infty$; (d) is obtained by applying Parseval's theorem to identify the effective noise term.

Assume that the BS knows the uplink transmit power $P_t$ and the noise variance $\sigma^2$, and let the channel gain be denoted as $E_{\mathbf{h}} = \|\mathbf{h}\|^2$, the effective received signal power as $E_{\mathbf{y}} = (\|\mathbf{y}\|^2 - N\sigma^2)/P_t$, and the transmit SNR as $\mathrm{SNR}_{\mathrm{tx}} = P_t/\sigma^2$. Given the well-known signal model in~\eqref{eqn:rx_power}, $E_{\mathbf{h}}$ can be estimated using the maximum likelihood approach~\cite{kay1993fundamentals}, by maximizing the likelihood function given by
\begin{align}
p(E_{\mathbf{y}};E_{\mathbf{h}}) 
= \frac{1}{\sqrt{2\pi\bigl(2 E_{\mathbf{h}}/\mathrm{SNR}_{\mathrm{tx}}\bigr)}} 
\exp\!\Bigl(-\tfrac{\bigl(E_{\mathbf{y}} - E_{\mathbf{h}}\bigr)^2}{2\bigl(2 E_{\mathbf{h}}/\mathrm{SNR}_{\mathrm{tx}}\bigr)}\Bigr).
\end{align}   Solving for $E_{\mathbf{h}}$ yields
\begin{align}
    \hat{E}_{\mathbf{h}} = -\frac{1}{\mathrm{SNR}_{\mathrm{tx}}} + \sqrt{\left(\frac{1}{\mathrm{SNR}_{\mathrm{tx}}}\right)^2 + E_{\mathbf{y}}^2}.
\end{align}
Then, substituting $E_{\mathbf{h}} \!=\! \left\|\mathbf{h}\right\|^2 \!=\! N\,h_0^2 \!=\! N\left(\frac{\lambda}{4\pi\,r_0}\right)^2$ from~\eqref{eqn:channel} into the above estimate leads to the following expression 
for the estimated range $r_0$
\begin{align}
    \hat{r}_0 = \min \left( r_{\text{max}},\ \max \left( r_{\text{min}},\ \frac{\lambda}{4\pi} \sqrt{\frac{N}{\hat{E}_{\mathbf{h}}}} \right) \right), \label{eqn:r_hat}
\end{align}
where $r_{\text{min}}$ and $r_{\text{max}}$ denote the minimum and maximum communication distances of the BS, respectively.

With $\hat{r}_0$ and \textbf{Definition}~\ref{def:signalspace}, the problem of estimating $\mathbf{V}_\epsilon$ is reduced to the following energy detection problem.
\vspace{-5pt}
\begin{align}
    \textbf{P1:} \quad & \max_{i} \quad \left\| \mathbf{y}_{i-g(i):i+g(i)} \right\|^2 = \sum_{\nu=i-g(i)}^{i+g(i)} |y_{\nu}|^2, \notag \\[5pt]
    \text{s.t.} \quad 
    & g(i) \!=\! L(\Delta_{(\theta_i, \hat{r}_0)}, \epsilon) 
    \!=\! \left\lfloor \sqrt{\frac{\Delta_{(\theta_i, \hat{r}_0)}^2}{2\pi} \left( \frac{\Delta_{(\theta_i, \hat{r}_0)}^2}{2\pi} \!+\! \frac{2}{\pi\epsilon} \right) } \right\rfloor, \tag{C1} \label{eqn:C1} \\[5pt]
    & \Delta_{(\theta_i, \hat{r}_0)} = N\sqrt{\frac{\pi d(1 - \theta_i^2)}{2\hat{r}_0}}, \tag{C2} \label{eqn:C2} \\[5pt]
    & \left| \left\| \mathbf{y}_{i-g(i):i-1} \right\|^2 
    - \left\| \mathbf{y}_{i+1:i+g(i)} \right\|^2 \right| < \eta. \tag{C3} \label{eqn:C3}
\end{align}
The constraints~\eqref{eqn:C1} and~\eqref{eqn:C2} are direct results of the analytical models presented in Subsection~\ref{subsec:3-b} and must be satisfied exactly, while the final constraint~\eqref{eqn:C3} ensures the symmetry of the $\epsilon$-approximated signal subspace, controlled by a small parameter $\eta$. \textbf{P1} can be efficiently implemented using the received signals within a sliding window framework, where the final constraint~\eqref{eqn:C3} is incorporated as a penalty function
\begin{align}
    \textbf{P2:} \quad & \max_{i} \quad \left\| \mathbf{y}_{i-g(i):i+g(i)} \right\|^2 \notag \\ 
    & \qquad\quad - \gamma \left| \left\| \mathbf{y}_{i-g(i):i-1} \right\|^2 
    - \left\| \mathbf{y}_{i+1:i+g(i)} \right\|^2 \right|, \notag \\[5pt]
    \text{s.t.} \quad 
    & \text{\eqref{eqn:C1} and~\eqref{eqn:C2}}. \notag
\end{align}
Here, $\gamma$ is a regularization parameter that penalizes asymmetry in the candidate solutions. Specifically, to estimate the angle of the UE while accounting for energy spread, the signal energy within a variable-length window---defined by~\eqref{eqn:C1} and~\eqref{eqn:C2}---is evaluated at each iteration as it slides across the sequence of received power measurements, $\left[ \left| y_1 \right|^2, \left| y_2 \right|^2, \dots, \left| y_N \right|^2 \right]$. The width of the sliding window at the $i$-th iteration models the degree of energy spread when the UE is located in the direction of $\theta_i$ with an estimated range given by~\eqref{eqn:r_hat}. After all evaluations, the window maximizing the objective function of \textbf{P2} is identified, and subsequently, the corresponding orthonormal Fourier basis $ \mathbf{\hat{V}}_\epsilon$ can be estimated as 
\begin{align}
    \Bigg[
        \mathbf{f}\left( \theta_{\hat{i}-g(\hat{i})} \right), \cdots , \mathbf{f}\left( \theta_{\hat{i}} \right), \cdots, \mathbf{f}\left( \theta_{\hat{i}+g(\hat{i})} \right)
    \Bigg],
    \label{eqn:hat_signalspace}
\end{align}
and the spatial angle $\theta_{0}$ can be estimated as $\bar{\theta}_{0} = \theta_{\hat{i}}$, where $\hat{i}$ is a solution of \textbf{P2}. In fact, the coarse alignment stage corresponds to energy detection while incorporating the energy spread based on the well-established model presented in Subsection~\ref{subsec:3-b}.

\begin{figure}[!t]
    \centering
    \includegraphics[width = 1\columnwidth]{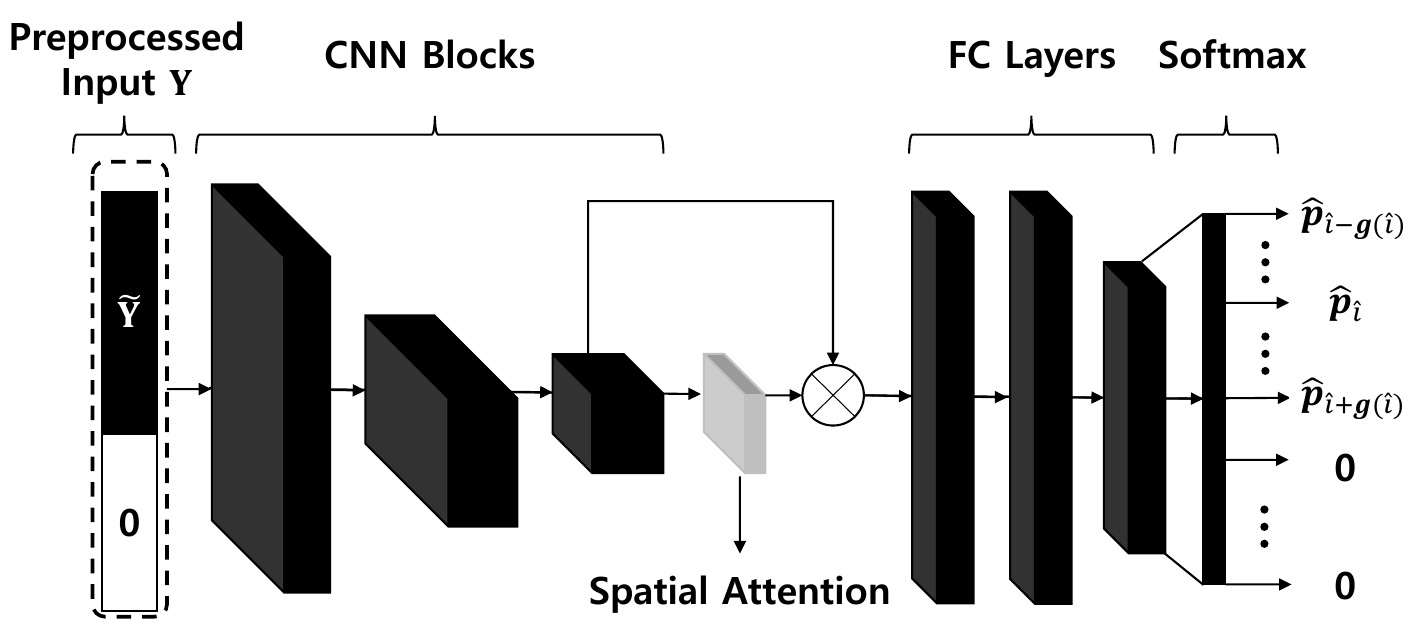}
    \caption{Proposed deep neural network structure.}
    \label{fig:NN}
\end{figure}

\subsubsection{Fine Alignment}
While $\theta_0$ and $r_0$ are estimated in the coarse alignment using an analytical framework, the downlink beam generated based on $(\bar{\theta}_0, \hat{r}_0)$ may experience a gain loss. This is primarily due to the discretized nature of the DFT matrix, as well as noise enhancement in low-power regimes. To address this issue, we propose a fine alignment stage employing a neural network that utilizes the estimated $\epsilon$-approximated signal subspace $\mathbf{V}_\epsilon$ to serve as a reduced search space. 

\paragraph{Input Layers}
Represent the vector of squared magnitudes of the received signals corresponding to $\mathbf{\hat{V}}_\epsilon$ by~\eqref{eqn:hat_signalspace} as
\begin{align}
    \mathbf{\tilde{Y}} = \left[ \left| y_{\hat{i}-g(\hat{i})} \right|^2, \dots, \left| y_{\hat{i}} \right|^2, \dots, \left| y_{\hat{i}+g(\hat{i})} \right|^2 \right]^T \in \mathbb{R}^{\left( 2g(\hat{i})+1 \right) \times 1}. \notag
\end{align}
To ensure a constant input dimension, we pad zeros to the end of $\mathbf{\tilde{Y}}$:
\begin{align}
    \mathbf{Y} = \Bigg[ \mathbf{\tilde{Y}}, \: \underset{U - \text{dim}(\mathbf{\tilde{Y}})}{\underbrace{0, \dots, 0}} \Bigg]^T \in \mathbb{R}^{U \times 1}, \label{eqn:input}
\end{align}
where $U = 2L(\Delta_{(\theta_0=0, r_0=r_{\text{min}})}, \epsilon) + 1$ represents the maximum energy spread, occurring when the UE is located at the position $(\theta_0, r_0)=(0, r_{\text{min}})$ with the most pronounced near-field effect. This input preprocessing implicitly separates the noise subspace from the signal subspace and leverages the neural network's inherent attention mechanism to focus on the relevant features, enhancing its ability to identify targets effectively. Using $\mathbf{Y}$ as input to the neural network, it is fed into the hidden layers. 

\paragraph{Convolutional Module}  
The convolutional module efficiently extracts features from the input data $\mathbf{Y}$~\eqref{eqn:input} through multiple convolutional blocks. Each block employs parallel convolutional layers with varying kernel sizes to capture multi-scale features, enhancing the network's ability to learn diverse patterns. The operation of each CNN block is defined by the following function,
\begin{align}
    \mathbf{O}_{\text{conv}}^{(w_{\text{cnn}})} \!=\! \Phi_{\text{PReLU}}^{(w_{\text{cnn}})} \! \left[ \!\Phi_{\text{Conv}_{1}}^{(w_{\text{cnn}})} \! \left( \! \mathbf{O}_{\text{conv}}^{(w_{\text{cnn}}-1)} \! \right) \! \oplus \! \dots \! \oplus \! \Phi_{\text{Conv}_{K}}^{(w_{\text{cnn}})} \! \left( \! \mathbf{O}_{\text{conv}}^{(w_{\text{cnn}}-1)} \! \right) \! \right] \!
\end{align}
where $\mathbf{O}_{\text{conv}}^{(w_{\text{cnn}})}$ represents the output feature maps from the $w_{\text{cnn}}$-th convolutional block ($w_{\text{cnn}} \in \{1, \dots, W_{\text{cnn}}\}$), $\Phi_{\text{Conv}_{k}}^{(w_{\text{cnn}})}$ denotes a convolution operation with kernel size $\kappa(k)$, and $\oplus$ indicates feature map concatenation. The function $\Phi_{\text{PReLU}}^{(w_{\text{cnn}})}(\cdot)$ represents the parametric rectified linear unit (PReLU), defined as
$\Phi_{\text{PReLU}}(x) = \max(0, x) + \delta \min(0, x),$ where $\delta$ is a learnable parameter controlling the slope for negative inputs, allowing the model to capture complex non-linear relationships.

\paragraph{Spatial Attention Module}  
The spatial attention module enhances feature representations by emphasizing important regions in the feature maps. Average-pooling and max-pooling are applied along the channel dimension to produce a two-channel representation, which is then processed by a 1D convolution. A sigmoid activation scales the resulting attention map to $[0, 1]$, and this attention map is applied to the feature maps via element-wise multiplication.  
\begin{align}
    \mathbf{O}_{\text{attn}} = \mathbf{O}_{\text{conv}}^{(W_{\text{CNN}})} \odot \Phi_{\text{sigmoid}} \big(\Phi_{\text{attnConv}} &\big(\Phi_{\text{AvgPool}}(\mathbf{O}_{\text{conv}}^{(W_{\text{CNN}})}) \notag \\
    &\; \oplus \Phi_{\text{MaxPool}}(\mathbf{O}_{\text{conv}}^{(W_{\text{CNN}})}) \big)\big)
\end{align}
where $\mathbf{O}_{\text{attn}}$ represents the feature maps after applying the attention mechanism, $\Phi_{\text{sigmoid}}$, $\Phi_{\text{attnConv}}$, $\Phi_{\text{AvgPool}}$, and $\Phi_{\text{MaxPool}}$ denote their respective operations, $\oplus$ indicates concatenation, and $\odot$ represents element-wise multiplication.

\paragraph{Global Average Pooling}  
After the spatial attention module, a global average pooling layer reduces the dimensionality of the refined feature maps, producing a compact representation suitable for the fully connected layers.  
\begin{align}
    \mathbf{O}_{\text{gap}} = \Phi_{\text{GAP}} (\mathbf{O}_{\text{attn}})
\end{align}
where $\Phi_{\text{GAP}}$ denotes the global average pooling operation.  

\paragraph{Fully Connected Layers}  
The fully connected (FC) layers map the extracted and refined features to the target output space.  Each FC layer is followed by a PReLU activation to enhance non-linear transformations. The operation of each FC block is defined by the following function, 
\begin{align}
    \mathbf{O}_{\text{fc}}^{(w_{\text{fc}})} = \Phi_{\text{PReLU}}^{(w_{\text{fc}})} \big( \mathbf{A}_{w_{\text{fc}}} \mathbf{O}_{\text{fc}}^{(w_{\text{fc}}-1)} + \mathbf{b}_{w_{\text{fc}}} \big)
\end{align}
where $\mathbf{O}_{\text{fc}}^{(w_{\text{fc}})}$ represents the output of the $w_{\text{fc}}$-th fully connected layer ($w_{\text{fc}} \in \{1, \dots, W_{\text{fc}}\}$), $\mathbf{A}_{w_{\text{fc}}}$ and $\mathbf{b}_{w_{\text{fc}}}$ correspond to the weight matrix and bias vector of the $w_{\text{fc}}$-th layer.  

\paragraph{Output Layers}  
The output layer estimates the true spatial angle within the reduced range $[\theta_{\hat{i}-g(\hat{i})}, \theta_{\hat{i}+g(\hat{i})}]$. To ensure valid predictions, a masking operation excludes invalid regions introduced by zero-padding in the input vector $\mathbf{Y}$. The masking operation assigns a value of negative infinity  $(-\infty)$ to invalid elements, effectively removing them from consideration during the softmax computation. After applying the mask and passing through the softmax layer, the probability distribution $\mathbf{\hat{p}}$ is given by
\begin{align}
    \mathbf{\hat{p}} &= \mathrm{softmax} (\mathbf{O}_{\text{fc}}^{(W_{\text{fc}})}) \notag \\
    &= [\hat{p}_{\hat{i}-g(\hat{i})}, \dots, \hat{p}_{\hat{i}}, \dots, \hat{p}_{\hat{i}+g(\hat{i})}, \underset{U - \text{dim}(\mathbf{\tilde{Y}})}{\underbrace{0, \dots, 0}}]^T
\end{align}
where $\hat{p}_{k}$ represents the probability assigned by the network to $k$ being the true spatial angle index. Using $\mathbf{\hat{p}}$ as a weight vector for  
\begin{align}
    \pmb{\vartheta} = [\theta_{\hat{i}-g(\hat{i})}, \dots, \theta_{\hat{i}}, \dots, \theta_{\hat{i}+g(\hat{i})}, \underset{U - \text{dim}(\mathbf{\tilde{Y}})}{\underbrace{0, \dots, 0}}]^T,
\end{align}
the refined spatial angle is computed as 
\begin{align}
    \hat{\theta}_0 = \mathbf{\hat{p}}^{T} \pmb{\vartheta}. \label{eqn:refined_theta}
\end{align}
During training, the model optimizes predictions using the cross-entropy loss function, defined as
\begin{align}
    J = -\sum_{j=\hat{i}-g(\hat{i})}^{\hat{i}+g(\hat{i})} p_{j} \log_{10} \hat{p}_{j},
\end{align}
where $p_{j} = 1$ if the target is $j$, and $p_{j} = 0$ otherwise.  

The entire neural network structure is shown in Fig.~\ref{fig:NN}. By narrowing the search space to the reduced space $\mathbf{\hat{V}}_\epsilon$, rather than evaluating all $N$ grids of the DFT matrix, the neural network focuses its attention on the relevant subspace. This approach improves computational efficiency, and more importantly, enhances the network's ability to accurately predict the spatial angle by leveraging the reduced candidate space. To summarize, by leveraging the domain knowledge of the signal subspace, the learning process becomes more efficient and robust, requiring fewer training samples and reducing computational complexity.

\begin{table}[!ht]
\centering
\caption{Proposed Neural Network Parameters.}
\label{tab:network_architecture_blocks}
{\footnotesize 
\renewcommand{\arraystretch}{1.75}
\begin{tabular}{|>{\centering\arraybackslash}p{0.14\linewidth}|>{\centering\arraybackslash}p{0.11\linewidth}|>{\centering\arraybackslash}p{0.60\linewidth}|}
\hline
\textbf{Module}        & \textbf{Block}            & \textbf{Parameter}                                               \\ \hline
\multirow{3}{*}{\parbox[m]{\linewidth}{\centering Convolution}} 
                        & Block 1                & \parbox[m]{18em}{\centering $C_{\text{in}}=1, C_{\text{out}}=16$, kernel sizes $=[3,5]$, Stride $=2$, BatchNorm, PReLU} \\ \cline{2-3}
                        & Block 2                & \parbox[m]{18em}{\centering $C_{\text{in}}=32, C_{\text{out}}=32$, kernel sizes $=[3,5]$, Stride $=2$, BatchNorm, PReLU} \\ \cline{2-3}
                        & Block 3               & \parbox[m]{18em}{\centering $C_{\text{in}}=64, C_{\text{out}}=64$, kernel sizes $=[3,5]$, Stride $=2$, BatchNorm, PReLU} \\ \hline
\multicolumn{2}{|c|}{Spatial Attention}        & $C_{\text{in}}=2, C_{\text{out}}=1$, kernel size $=7$, Sigmoid                        \\ \hline
\multicolumn{2}{|c|}{Global Pooling}               & Adaptive Average Pooling                                         \\ \hline
\multirow{3}{*}{\parbox{\linewidth}{\vspace{0cm} \centering Fully Connected}} 
                        & Block 1                  & $F_{\text{in}}=128, F_{\text{out}}=128$, BatchNorm, PReLU        \\ \cline{2-3}
                        & Block 2                  & $F_{\text{in}}=128, F_{\text{out}}=128$, BatchNorm, PReLU        \\ \cline{2-3}
                        & Output           & $F_{\text{in}}=128, F_{\text{out}}=U$                           \\ \hline
\multicolumn{2}{|c|}{Dropout}                      & $p=0.5$                                                          \\ \hline
\end{tabular}
}
\end{table}

\section{Simulation Results}\label{sec:simulation}

In this section, we evaluate the performance of the proposed beam alignment scheme alongside benchmark schemes from prior works, with Monte Carlo simulations averaged over 50,000 iterations. For the simulation setup, the BS employs a half-wavelength spaced ULA with $N=256$ antennas, operating at $28$ GHz with $850$ MHz bandwidth~\cite{ali2017OOB}. The noise power spectral density is $N_0=-174$ dBm/Hz, yielding a noise power of $\sigma^2 \approx -84.7058$ dBm. The physical angle of the UE is uniformly distributed in $\phi_0 \sim \mathcal{U}(-\pi/3, \pi/3)$, and the distance between the BS and UE is uniformly distributed in $r_0 \sim \mathcal{U} (4 \, \text{m}, 80 \, \text{m})$. This distance lies in the radiative near-field, bounded by the Fresnel distance $d_F \approx 3.44 \, \text{m}$ and the Rayleigh distance $d_R \approx 348.35 \, \text{m}$.

The performance of the proposed scheme, which consists of two stages, is shown in two steps: the proposed scheme with only the coarse alignment stage and the proposed scheme with the additional fine alignment stage. In our simulation setup, we set the parameters $\epsilon=0.1$ and $\gamma=P_t/10^{1.5}$. In this case, the maximum energy spread is given by $U = 2L(\Delta_{(\theta_0=0, r_0=r_{\text{min}})}, \epsilon) + 1 = 49$. For  the proposed scheme with only the coarse alignment stage, $\bar{\theta}_{0}$, the solution of \textbf{P2}, and $\hat{r}_0$ in~\eqref{eqn:r_hat} are used to construct the optimized beam through~\eqref{eqn:nf_manifold}. For  the proposed scheme with the additional fine alignment stage, the refined $\hat{\theta}_0$ obtained through~\eqref{eqn:refined_theta} and $\hat{r}_0$ in~\eqref{eqn:r_hat} are used to construct the optimized beam. The neural network architecture utilized is detailed in Table~\ref{tab:network_architecture_blocks}. We employ the Adam optimizer with an initial learning rate of $0.001$ and a cosine annealing learning rate scheduler for gradual adjustment. The training process was regulated by early stopping based on validation performance, with a maximum cap of 100 epochs.

For comparison, we evaluate the following benchmark schemes:

\begin{enumerate}
    \item \textbf{Least squares (LS):} The channel vector $\mathbf{h}$ is directly estimated from~\eqref{eqn:stacked_rx_signal} using the least squares method as $\hat{\mathbf{h}} = \mathbf{F}\mathbf{y}/\sqrt{P_t}$. Here we use $(\mathbf{F}^H)^{-1} = \mathbf{F}$, utilizing the unitary property of the DFT matrix $\mathbf{F}$ in~\eqref{eqn:DFT}.

    \item \textbf{Polar codebook-based exhaustive search (Polar-Exh)~\cite{cui2023rainbow}:} All codewords in the polar-domain codebook, as defined in~\eqref{eqn:polar_trans}, are tested to identify the codeword $\mathbf{a}(\theta_{p}, r_{p,q})$ that maximizes energy in its direction. The optimal indices are determined by $(n^{*}, q^{*}) = \argmax_{n, q} \left| \sqrt{P_t} {\mathbf{a}(\theta_n, r_{n,q})}^H \mathbf{h} + \mathbf{z}_{n,q} \right|$.

    \item \textbf{Angle Support Width based Joint Angle and Range Estimation (ASW-JE)~\cite{you2024dft}:} The spatial angle $\theta_{\text{Median}(\mathcal{B})}$ is estimated from the index set of codewords $\mathcal{B}$, which satisfy sufficient received power; $\mathcal{B} = \left\{i \;\middle|\; \left| y_i \right| > \kappa^2 \max_{\mathbf{f}(\theta_i)} \left| y_i \right| \right\},$ where $y_i$ is defined in~\eqref{eqn:rx_signal}.  The range is estimated by demapping the pre-derived approximation of the noiseless correlation function with the ratio $\left| y_i \right| / \left| y_{\text{Median}(\mathcal{B})} \right|$, using a numerical line search on $\mathcal{Z} = \left\{ \varpi = \sqrt{\frac{r}{d(1-\theta^2)}} \,\middle|\, \varpi = \varpi_{\text{min}}, \varpi_{\text{min}} + \Delta \varpi, \dots, \varpi_{\text{max}} \right\}.$ with step size $\Delta \varpi = 0.1.$

    \item \textbf{DNN-aided reduced DFT codebook-based beam training (DFT-DNN)~\cite{wang2023naivecnn}:} $N/4$-point DFT beams are tested, and the received signal is mapped to a polar-domain codeword $\mathbf{a}(\theta_{n^{*}}, r_{n^{*},q^{*}})$ using two DNNs, $G_{\text{angle}}(\cdot)$ and $G_{\text{range}}(\cdot)$. The optimal angle and range indices are given by $n^{*} = \argmax_{n} [G_{\text{angle}}(\mathbf{y}_{N/4\text{-DFT}})]_{n}$ and $q^{*} = \argmax_{q} [G_{\text{range}}(\mathbf{y}_{N/4\text{-DFT}})]_{q}$, where $\mathbf{y}_{N/4\text{-DFT}}$ represents the received signal vector processed by the $N/4$-point DFT matrix.

    \item \textbf{Deep learning-based near-field beam training (DNBT)~\cite{qi2023dnbt}:} Uniformly selected codewords $\mathbf{a}(\theta_{n(q)}, r_{n(q),q})$ are tested, where $n(q) = \left( \left( n_0(q) - 1 + k\chi \right) \:\mathrm{mod}\: N \right) + 1, \quad k = 1, \dots, \lfloor N/\chi \rfloor,$ are sampled from the polar-domain codebook in~\eqref{eqn:polar_trans}. The received signal is then mapped to a polar-domain codeword $\mathbf{a}(\theta_{n^{*}}, r_{n^{*},q^{*}})$ using a DNN, $G_{\text{DNBT}}(\cdot)$, with optimal indices computed as $(n^{*}, q^{*}) = \argmax_{n, q} [G_{\text{DNBT}}(\mathbf{y}_{\text{(DNBT)}})]_{n,q}.$ We set $\chi = 16$ to maintain the same compression ratio in the angular-domain as in~\cite{qi2023dnbt}, with $n_0(q) = 1 + 2(q - 1)$.
\end{enumerate}

For the polar-domain codebook-based schemes---Polar-Exh~\cite{cui2023rainbow}, DFT-DNN~\cite{wang2023naivecnn}, and DNBT~\cite{qi2023dnbt}---, we set the column coherence parameter $\beta_\Delta = 1.2$, as suggested in~\cite{dai2022nfce}. This configuration results in a polar-domain codebook size of $NQ = 256 \times 16 = 4096$, which is 16 times larger than the $N$-point DFT codebook size of $N = 256$. Furthermore, the benchmark schemes---ASW-JE~\cite{you2024dft}, DFT-DNN~\cite{wang2023naivecnn}, and DNBT~\cite{qi2023dnbt}---are evaluated alongside their improved versions from each respective paper. For ASW-JE~\cite{you2024dft}, $K_{\mathrm{a}} = 3$ angle-range pairs are estimated, and the evaluation involves testing additional polar-domain codewords corresponding to each pair. For DFT-DNN~\cite{wang2023naivecnn} and DNBT~\cite{qi2023dnbt}, an additional $K_{\mathrm{b}} = 20$ polar-domain codewords are tested, corresponding to the top $K_{\mathrm{polar}}$ most probable candidates inferred by the respective models. Their neural networks are reproduced as described in the respective papers, as they operate on the same polar grid size of $NQ = 4096$.

\begin{figure}[!t]
    \centering
    \subfigure[]{
        \includegraphics[width = 1\columnwidth] {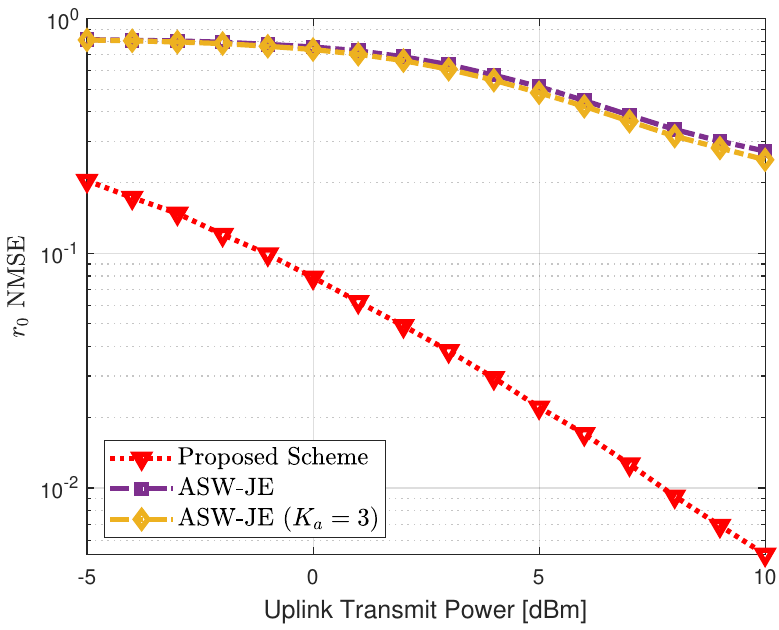}
    \label{fig:range_NMSE}
    }
    \subfigure[]{
        \includegraphics[width = 1\columnwidth]{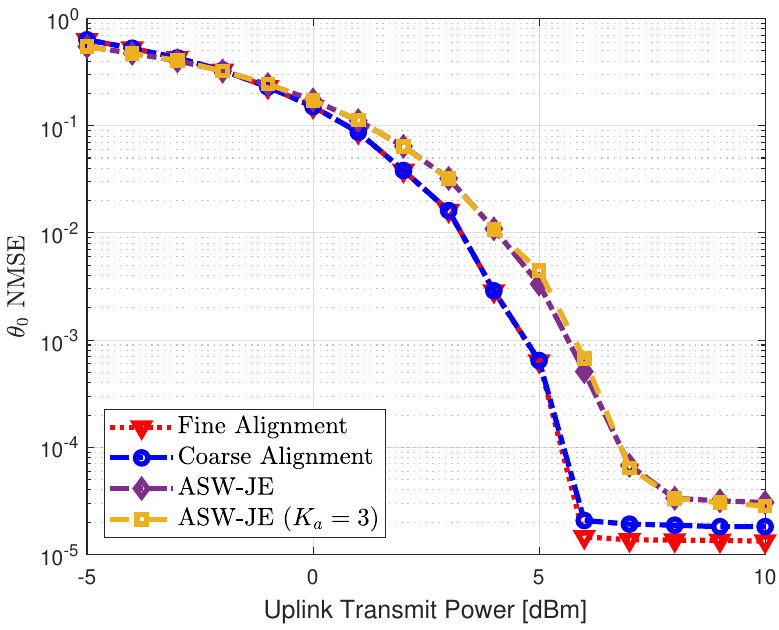}
        \label{fig:theta_NMSE}
    }
    \caption{NMSE versus transmit power for (a) range $r_0$ and (b) spatial angle $\theta_0.$}
    \label{fig:NMSE}
\end{figure}

\subsection{Beam Alignment Performance} \label{subsec:BA_performance}
First, we evaluate the normalized mean squared error (NMSE) performance for range estimation of $r_0$ and spatial angle estimation of $\theta_0$, as shown in Fig.~\ref{fig:NMSE}. NMSE is defined as:
\begin{align*}
    \mathrm{NMSE}_{\mathrm{range}} &:= \frac{\mathbb{E}\left\{ \left| r_0 - r_{\mathrm{est}} \right|^2 \right\}}{\mathbb{E}\left\{ \left| r_0 \right|^2 \right\}}, \\
    \mathrm{NMSE}_{\mathrm{angle}} &:= \frac{\mathbb{E}\left\{ \left| \theta_0 - \theta_{\mathrm{est}} \right|^2 \right\}}{\mathbb{E}\left\{ \left| \theta_0 \right|^2 \right\}}.
\end{align*}
Since LS, Polar-Exh~\cite{cui2023rainbow}, DFT-DNN~\cite{wang2023naivecnn}, and DNBT~\cite{qi2023dnbt} do not directly estimate $r_0$ and $\theta_0$, we compare the proposed scheme with only the ASW-JE~\cite{you2024dft} and its improved variant, ASW-JE ($K_a=3$).

The $\mathrm{NMSE}_{\mathrm{range}}$ performance is illustrated in Fig.~\ref{fig:range_NMSE}. The proposed scheme effectively exploits the received signal power (gain) and the unitary properties of the DFT matrix, as formulated in~\eqref{eqn:rx_power}, achieving superior performance over ASW-JE\cite{you2024dft}. In contrast, ASW-JE relies on a numerical search based on the ratio of noisy measurements and exhibits a more rapid performance degradation as SNR increases. Our simulation setting encompasses a broad range of $r_0$ values, making it more practical but also resulting in a lower average received SNR. This, in turn, further degrades the performance of ASW-JE compared to the results presented in~\cite{you2024dft}.

\begin{figure}[!t]
    \centering
    \includegraphics[width = 1\columnwidth]{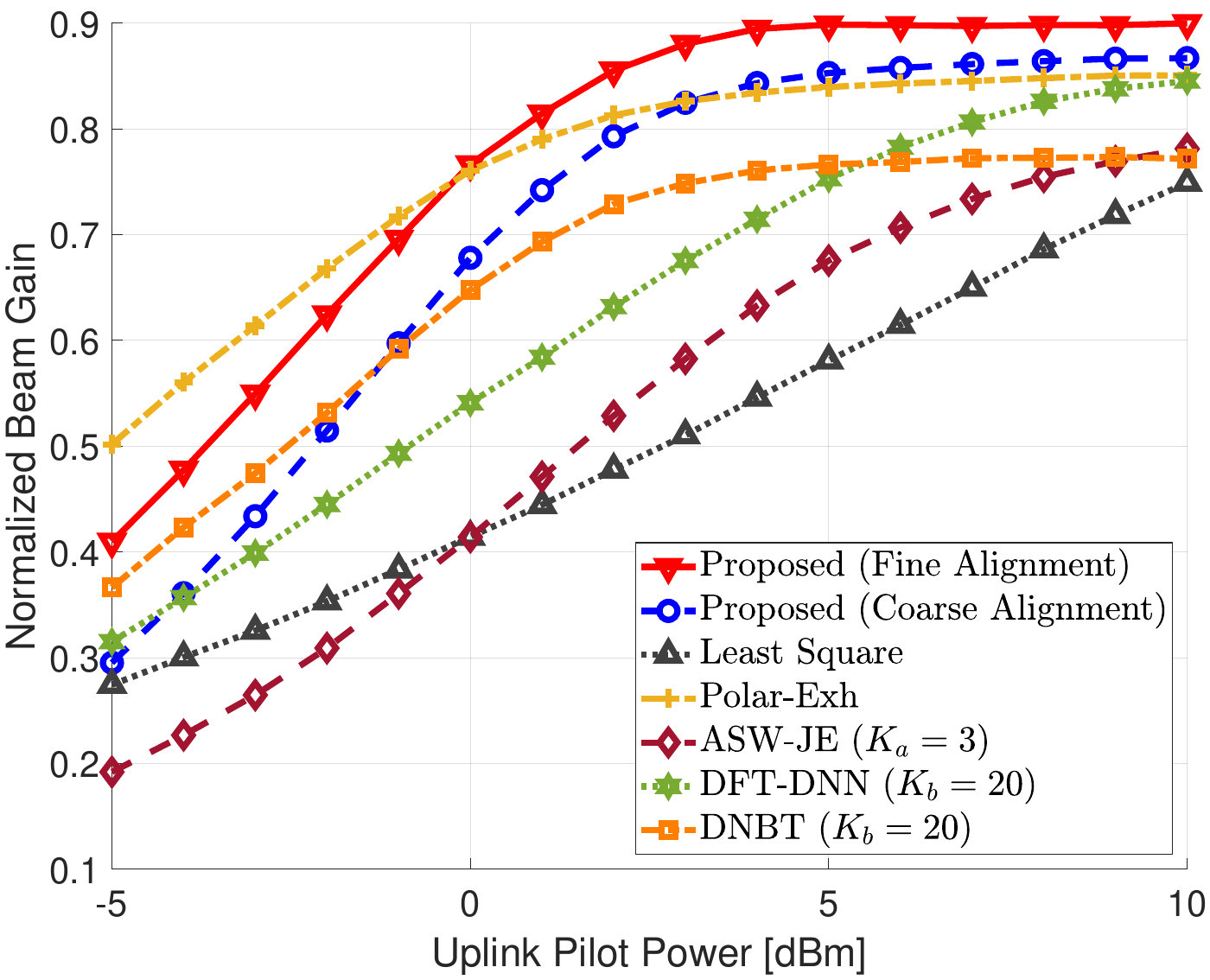}
    \caption{Normalized beam gain versus transmit power.}
    \label{fig:beamgain}
\end{figure}

In Fig.~\ref{fig:theta_NMSE}, we evaluate $\mathrm{NMSE}_{\mathrm{range}}$ performance with and without the fine alignment stage, along with ASW-JE~\cite{you2024dft}. Since our approach explicitly models the energy spread effect and incorporates it into both stages, it outperforms ASW-JE, which relies solely on empirical observations. The performance gap between with and without the fine alignment stage remains negligible for transmit power levels below $6\;\text{dBm}$ due to errors in estimating an $\epsilon$-approximated signal subspace during the coarse alignment stage. Particularly, when the estimated $\epsilon$-approximated signal subspace~\eqref{eqn:hat_signalspace} \textit{does not} include the DFT column corresponding to the direction of the true UE's spatial angle, fine alignment stage, while refining the estimation within a reduced search space, fails to further improve the estimation. Furthermore, since such errors have a more significant impact on NMSE  than the beam gains provided by fine alignment stage, the NMSE plot alone cannot fully capture the effectiveness of the proposed fine alignment stage. Indeed, further  simulations to be explained below indicate that when the $\epsilon$-approximated subspace is accurately estimated during coarse alignment stage, additional fine alignment stage significantly enhances the estimation across the entire power range. The small performance gap between with and without the fine alignment stage results from approximation errors in~\eqref{eqn:correlation}, which are mitigated by the DNN-aided fine alignment stage.
%These results demonstrate that the proposed scheme can be effectively applied to parameter estimation, which is often assumed to be perfectly known in antenna configuration designs~\cite{yun2024analogdigital, palaiologos2024losnonuniform}. 

To clearly observe the gain offered by the proposed scheme, we evaluate the normalized beam gain, defined as $\left| \mathbf{w}_{\mathrm{opt}}^H \mathbf{h} \right|/ \left\| \mathbf{h} \right\|$, where $\mathbf{w}_{\mathrm{opt}}$ represents the optimized beam for each scheme. As shown in Fig.~\ref{fig:beamgain}, even in the low-power regime, the proposed scheme with the additional fine alignment stage achieves slightly lower performance compared to the exhaustive search, Polar-Exh~\cite{cui2023rainbow}, but exceeds that of the other baselines. As the transmit power increases, the proposed scheme maintains its advantage over benchmark schemes and outperforms Polar-Exh~\cite{cui2023rainbow}. This improvement results from the effective off-grid estimation of $r_0$, as given in~\eqref{eqn:r_hat} and illustrated in Fig.~\ref{fig:range_NMSE}, leveraging the unitary property of the DFT matrix. In contrast, polar-domain codebook-based approaches such as Polar-Exh~\cite{cui2023rainbow}, DFT-DNN~\cite{wang2023naivecnn}, and DNBT~\cite{qi2023dnbt} rely on a predefined range grid, limiting their flexibility in range estimation. Notably,  even the proposed scheme with only the coarse alignment stage demonstrates superior performance in the high-power regime, outperforming the Polar-Exh~\cite{cui2023rainbow} and achieving a level comparable to the proposed scheme with the fine alignment stage. 

Moreover, by comparing  Fig.~\ref{fig:theta_NMSE} and Fig.~\ref{fig:beamgain}, we observe that incorporating the fine alignment stage yields a particularly significant improvement in average beam gain  (Fig.~\ref{fig:beamgain}), even though the average NMSE improvement remains marginal in the low-power regime (Fig.~\ref{fig:theta_NMSE}). This phenomenon is primarily attributed to the narrow beam characteristics of the XL-MIMO system. Specifically, when the estimated $\epsilon$-approximated signal subspace~\eqref{eqn:hat_signalspace} includes the DFT column corresponding to the true UE's spatial angle, the fine alignment stage refines the estimation. This refinement results in a more precisely directed beam, substantially enhancing beam gain in the low-power regime. However, NMSE performance remains constrained in this regime because estimation errors in identifying the 
 $\epsilon$-approximated signal subspace have a dominant impact on the \textit{average} metric, i.e., NMSE. Despite this, the beam gain analysis confirms the fine alignment stage's effectiveness even under low-power conditions. In the high-power regime, although the NMSE performance gap between the two-stage approach is relatively small, the beam gain improvement remains pronounced due to the high beamforming gain by an accurate and narrow beam. Overall, beam gain improvement significantly exceeds NMSE improvement, primarily due to the system’s narrow beam characteristics. Analyzing these two performance metrics underscores both the necessity and effectiveness of the fine alignment stage.

\begin{figure}[!t]
    \centering
    \includegraphics[width = 1\columnwidth]{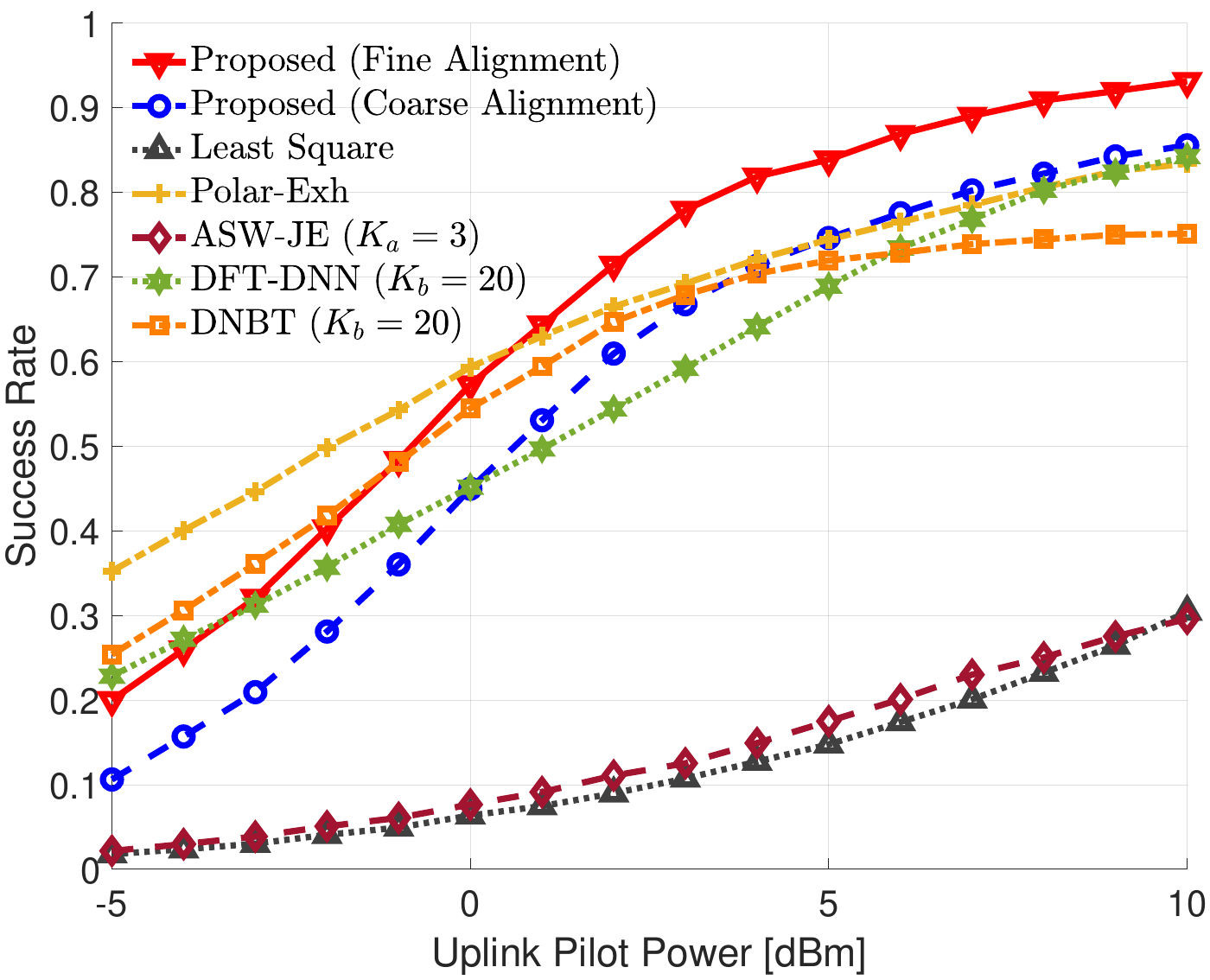}
    \caption{Success rate versus transmit power.}
    \label{fig:successrate}
\end{figure}

In Fig.~\ref{fig:successrate}, we plot the success rate, which is defined as the ratio of successful trials to the total number of trials in the Monte Carlo simulation. A trial is considered successful if, for a given range, angle, and corresponding channel $\mathbf{h}$ in~\eqref{eqn:channel}, the optimized beam gain is greater than or equal to the beam gain achieved by the optimal polar-domain codeword, which is genie-aided and represents the best possible selection within the polar-domain codebook. That is, the beam gain obtained from the corresponding quantized polar-domain codeword for the given range and angle. Mathematically, a trial is deemed successful if  
\begin{equation}
    \mathrm{success} := \left| \frac{\mathbf{w}_{\mathrm{opt}}^H \mathbf{h}}{\left\| \mathbf{h} \right\|} \right| \geq \max_{\mathbf{a}(\theta_m, r_{m,q}) \in  \mathbf{F}_{\text{Polar}}~\eqref{eqn:polar_trans}} \left| \frac{\mathbf{a}(\theta_m, r_{m,q})^H \mathbf{h}}{\left\| \mathbf{h} \right\|} \right|.
\end{equation}
The success rate is given by  
\begin{equation}
    P_{\text{success}} = \frac{1}{\mathcal{T}} \sum_{t=1}^{\mathcal{T}} \mathbb{I} \left( \left| \frac{\mathbf{w}_{(t),\mathrm{opt}}^H \mathbf{h}_{(t)}}{\left\| \mathbf{h}_{(t)} \right\|} \right| \geq \max_{\mathbf{a} \in  \mathbf{F}_{\text{Polar}}~\eqref{eqn:polar_trans}} \left| \frac{\mathbf{a}^H \mathbf{h}_{(t)}}{\left\| \mathbf{h}_{(t)} \right\|} \right| \right),
\end{equation}
where $ \mathbf{h}_{(t)} $ denotes the channel vector at the $ t $-th trial, $ \mathbf{w}_{(t),\mathrm{opt}} $ is the optimized beam for a given scheme at the $ t $-th trial, $ \mathcal{T} $ is the total number of Monte Carlo trials, and $ \mathbb{I}(\cdot) $ is the indicator function, which equals 1 if the condition inside holds and 0 otherwise. Note that as $ \mathcal{T} $ increases, the empirical success rate converges to the true \textit{success probability} by the law of large numbers.
As shown in Fig.~\ref{fig:successrate}, both proposed schemes perform exceptionally well in the high-power regime. In particular, the proposed scheme with the fine alignment stage demonstrates robust performance after $P_t \approx 0 \;\text{dBm}$. Moreover, even the proposed scheme with only the coarse alignment stage outperforms the benchmarks after $P_t\approx 4\;\text{dBm}.$

\begin{figure}[!t]
    \centering
    \subfigure[]{
        \includegraphics[width = 1\columnwidth]{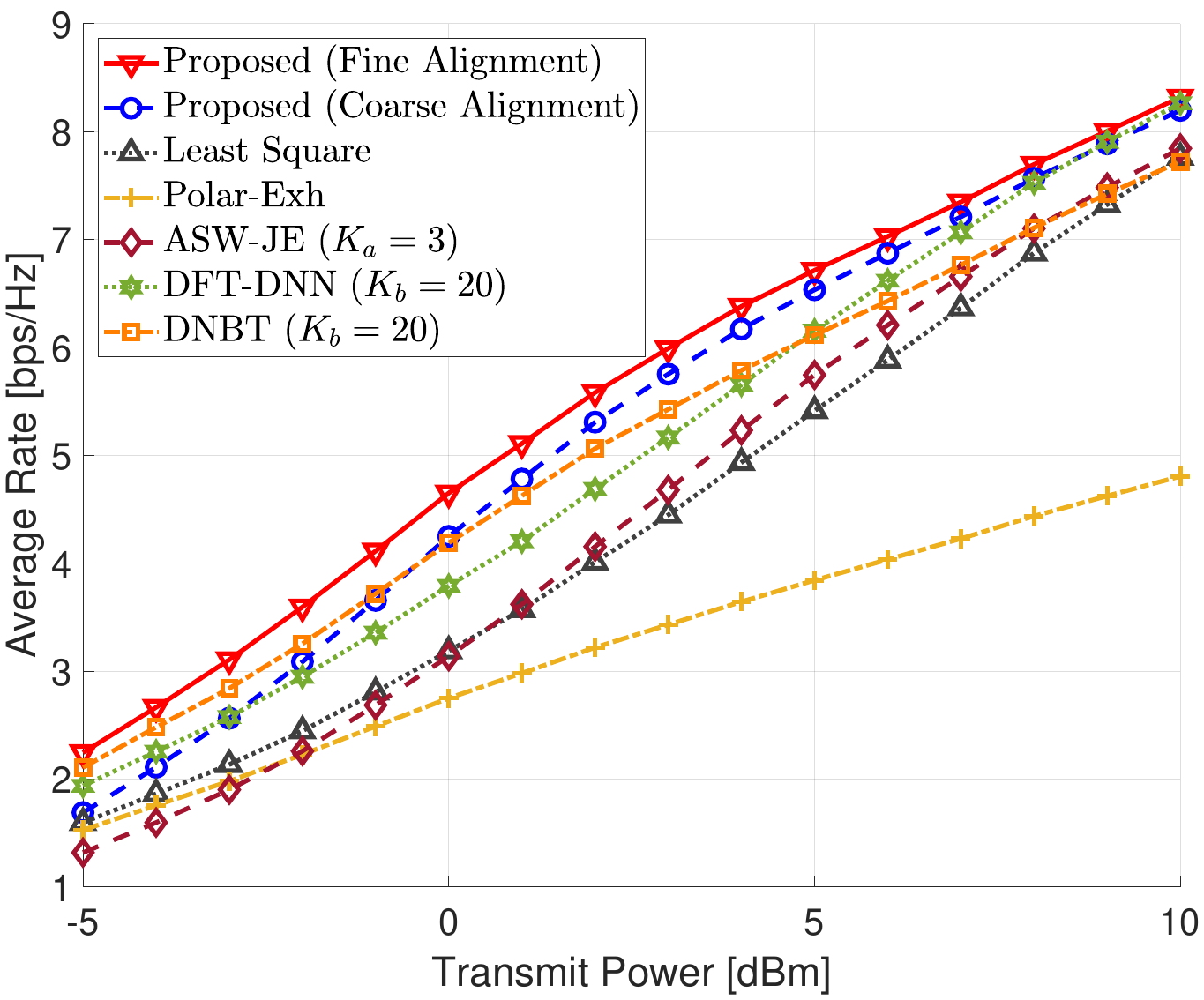}
    \label{fig:rate_1}
    }
    \subfigure[]{
        \includegraphics[width = 1\columnwidth]{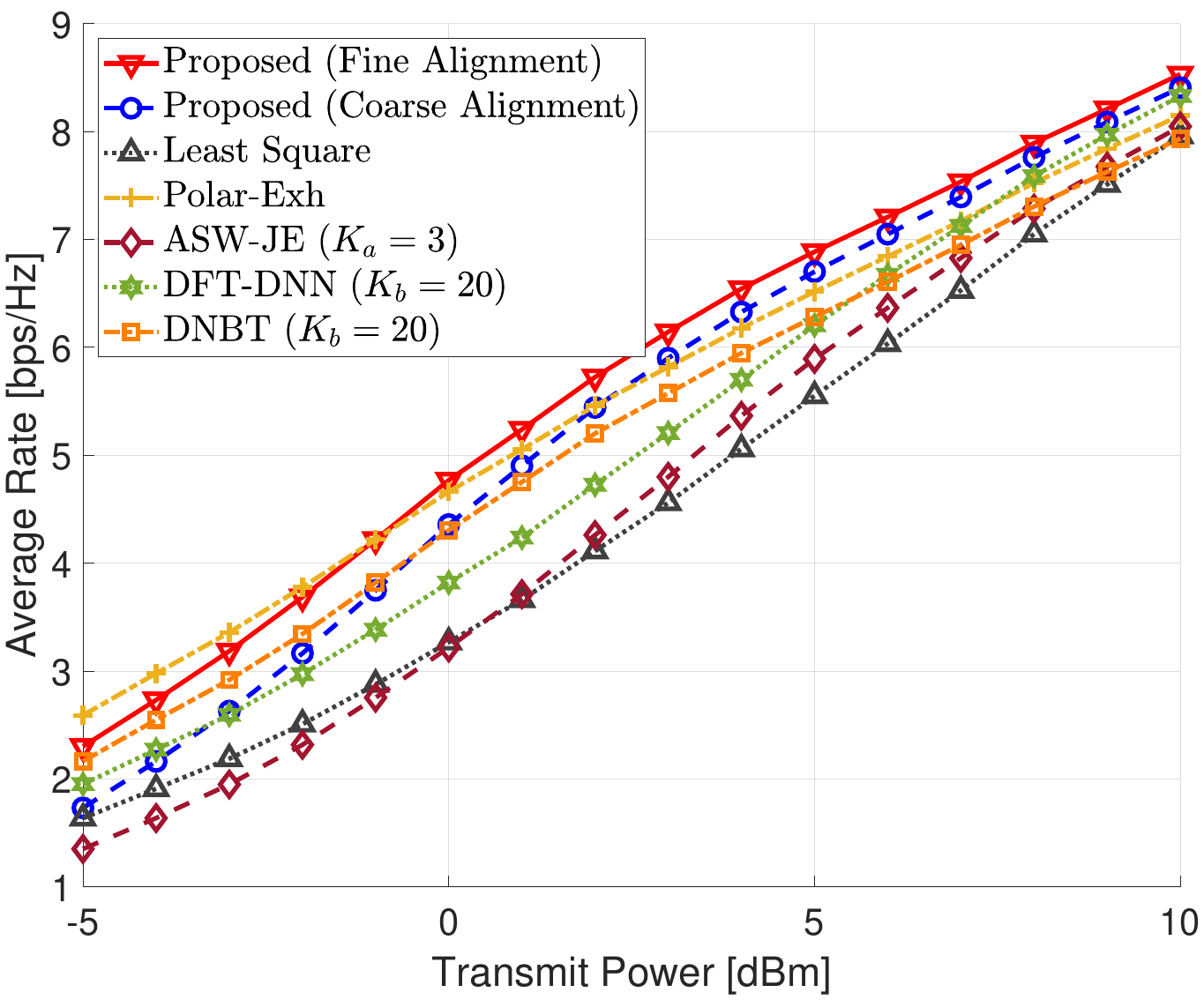}
        \label{fig:rate_16}
    }
    \caption{Normalized achievable rate versus transmit power for (a) $1$ RF-chain and (b) $16$ RF-chain at the BS.}
    \label{fig:achievable_rate}
\end{figure}

In Fig.~\ref{fig:achievable_rate}, we evaluate the achievable rate normalized by the time consumed for beam alignment, given by $R = \left( 1 - \frac{T_{\text{ba}}}{T_{\text{total}}} \right) \log_2 \left( 1 + \frac{P_t \left| \mathbf{w}^H_{\text{opt}} \mathbf{h} \right|}{\sigma^2} \right),$
where $T_{\text{ba}}$ denotes the time allocated for beam alignment, and $T_{\text{total}}$ represents the total time available for both beam alignment and data transmission, satisfying $T_{\text{total}} \geq T_{\text{ba}}$. As in \cite{kang2024pilot}, the time slot for each measurement is set to $T_{\text{symbol}} = 1.04 \;\mu\text{s}$, corresponding to the symbol duration for a subcarrier spacing of $960\;\text{kHz}$. This subcarrier spacing is expected to be utilized in 6G systems employing sub-terahertz communications. The total frame duration, $T_{\text{total}}$, is set to $10 \;\text{ms}$, which matches the frame length in 5G. The corresponding symbol times required for each scheme are listed in Table~\ref{tab:symboltimes}.

\begin{table}[!ht]
\centering
\caption{Time for the beam alignment.}
\label{tab:symboltimes}
\begin{tabular}{|c|c|c|}
\hline
\textbf{Scheme} & \textbf{Number of Symbols} & \textbf{BA Time} \\ \hline
Proposed (Both) & $N=256$ & $256 T_{\text{symbol}}$ \\ \hline
LS& $N=256$ & $256 T_{\text{symbol}}$ \\ \hline
Polar-Exh~\cite{cui2023rainbow} & $NQ=4096$ & $4096 T_{\text{symbol}}$ \\ \hline
ASW-JE $K_a=3$~\cite{you2024dft} & $N=259$ & $259 T_{\text{symbol}}$ \\ \hline
DFT-DNN $K_b=20$~~\cite{wang2023naivecnn} & $N/4 + K_b=84$ & $84 T_{\text{symbol}}$ \\ \hline
DNBT $K_b=20$~\cite{qi2023dnbt}  & $\lfloor N/\chi \rfloor)Q + K_b=276$ & $276 T_{\text{symbol}}$\\ \hline 
\end{tabular}

\end{table}

For convenience, we assume equal power allocation for both pilot and data transmission. We evaluate the normalized achievable rate performance for a BS with $N_{\text{RF}}=1$ RF-chain analog beamforming and $N_{\text{RF}}=16$ RF-chain fully connected analog-digital hybrid systems. The analysis remains the same for both cases, with the only difference being the number of measurements taken at the BS, which is $1$ and $16$, respectively. Consequently, for the $N_{\text{RF}}=16$ RF-chain fully connected analog-digital hybrid system, the number of pilot symbols required for beam alignment and the corresponding alignment time are computed as $\lceil \text{value in Table~\ref{tab:symboltimes}} / 16 \rceil$.

In Fig.~\ref{fig:rate_1}, the normalized achievable rate for $N_{\text{RF}}=1$ is presented. As shown in the figure, the proposed scheme with the additional  fine alignment stage consistently achieves superior performance across the entire regime, while the proposed scheme with only the coarse alignment stage demonstrates competitive performance as the transmit power increases. Specifically, after $-1\;\text{dBm}$,  the latter outperforms the benchmark schemes as well. Notably, the excessive beam alignment time required by Polar-Exh~\cite{cui2023rainbow} significantly degrades its performance. 

However, in Fig.~\ref{fig:rate_16}, as the beam alignment time decreases due to the reduced number of pilot symbols in $N_{\text{RF}}=16$ systems, Polar-Exh~\cite{cui2023rainbow} achieves the best performance in the very low-power regime. Nevertheless, as the transmit power increases, the proposed scheme outperforms all benchmark methods.

Lastly, from Figs.~\ref{fig:beamgain}, ~\ref{fig:successrate}, and ~\ref{fig:achievable_rate}, we conclude that the fine alignment stage effectively enhances beam alignment performance, as previously discussed, since the three performance evaluations consistently demonstrate its superiority over the coarse alignment stage across the entire power regime.  Furthermore, although DFT-DNN~\cite{wang2023naivecnn} and DNBT~\cite{qi2023dnbt} employ large neural networks to process the entire set of received signal measurements for optimal beam estimation—as analyzed in the following subsection—they offer no performance advantage over the proposed scheme, even in the low-power regime. In contrast, the proposed two-stage approach enhances efficiency by incorporating an additional preprocessing step of the coarse alignment stage to extract the relevant space for the fine alignment stage. This structured design results in superior beam alignment performance, particularly in low-power scenarios.

\subsection{Computational Complexity} \label{subsec:complexity}
Another important aspect of performance evaluation is computational complexity. To ensure a fair comparison between the proposed scheme and all benchmark schemes, both with and without neural networks, we calculate the number of floating-point operations (FLOPs). Specifically, each addition, multiplication, and division is counted as one FLOP. The same assumption is applied to operations such as square roots, minimum, maximum, and absolute value calculations. Additionally, we disregard vectorization or parallel processing, so the results represent a pure operation count.
 
For the coarse alignment stage of proposed scheme, the overall computation is dominated by the addition involved in range estimation~\eqref{eqn:rx_power} and sliding window operation for the estimating approximated signal subspace in \textbf{P2}. However, if we use the cumulative summation which calculates the running total of a sequence in~\eqref{eqn:rx_power}, the computation redundancy is significantly reduced. In that case, there are a total of
\begin{align}
    F_{\text{coarse}} &\!=\! 
    \underset{\text{\eqref{eqn:rx_power}}}{\underbrace{3N}} \!+\! 
    \underset{\text{cumulative sum}}{\underbrace{N}} \!+\! 
    \underset{\text{\eqref{eqn:C1} \&~\eqref{eqn:C2}}}{\underbrace{5N}} \!+\!  
    \underset{\text{sliding window loop}}{\underbrace{8N}} \!+\!
    7 \notag \\ 
    &= 17N+7
\end{align}
FLOPs, where $7$ accounts for all the computations irrelevant to the loop operations. For fine alignment, we calculate the FLOPs for each DNN layer as follows~\cite{lee2023flop}
\begin{align}
    F(\ell) =
    \begin{cases} 
    (2 C(\ell') K(\ell)^2 \!-\! 1) H(\ell) W(\ell) C(\ell), \!&\! \text{if } \tau(\ell) \!=\! \text{CNN}, \\
    K(\ell)^2 H(\ell) W(\ell) C(\ell), \!&\! \text{if } \tau(\ell) \!=\! \text{Pool}, \\
    (2 \xi(\ell') \!-\! 1) \xi(\ell), \!&\! \text{if } \tau(\ell) \!=\! \text{FC}.
    \end{cases} \label{eqn:FLOP}
\end{align}
Here, $\ell$ and $\ell' = \ell - 1$ are the indices of the current and preceding layers, respectively. $\tau(\ell)$ denotes the type of the $\ell$-th layer, where $\tau(\ell) \in \{\text{CNN}, \text{Pool}, \text{FC}\}$. $H(\ell), W(\ell), C(\ell), K(\ell)$ represent the height, width, number of channels, and kernel size at the $\ell$-th layer, respectively, while $\xi(\ell)$ is the number of neurons in the $\ell$-th layer. Using~\eqref{eqn:FLOP} and network structure in TABLE~\ref{tab:network_architecture_blocks}, the total number of FLOPs is
\begin{align}
    F_{\text{Fine}} \!=\! \sum_{\ell} F(\ell) &\!=\!
    \underset{\text{CNN Block 1}}{\underbrace{224L_1}} \!+\! 
    \underset{\text{CNN Block 2}}{\underbrace{16,320L_2}} \!+\! 
    \underset{\text{CNN Block 3}}{\underbrace{65,408L_3}} \notag \\  
    &\quad \!+\! \underset{\text{Spatial Attention}}{\underbrace{155L_3}} 
     \!+\! \underset{\text{Pooling}}{\underbrace{128L_3}} \notag \\  
    &\quad \!+\!
    \underset{\text{FC Block 1}}{\underbrace{32,640}} \!+\!
    \underset{\text{FC Block 2}}{\underbrace{32,640}} \!+\!
    \underset{\text{FC Block 3}}{\underbrace{255U}}
\end{align}
where $L_1=\lfloor (U+1)/2 \rfloor$, $L_2=\lfloor (L_1 + 1)/2 \rfloor$, $L_3=\lfloor (L_2 + 1)/2 \rfloor$ and $U$ is the fixed input feature length, maximum energy spread. Therefore, the total FLOPs of proposed scheme is 
\begin{align}
    F_{\text{total}} &= F_{\text{coarse}} + F_{\text{fine}} \notag \\
    &\approx (17N + 7) + (12,658U+65,280).
\end{align}
In a similar way, the FLOPs of various benchmarks can be computed.

\begin{enumerate}
    \item 
    \textbf{LS:} The computation involves multiplying an $N \times N$ complex matrix with an $N \times 1$ complex column vector. The FLOPs can be expressed as $F_{\text{LS}} = 8N^2 - 2N.$

    \item 
    \textbf{Polar-Exh~\cite{cui2023rainbow}:} The main computational cost comes from computing the power of the received signal and comparing values to find the maximum. This results in $F_{\text{Polar-Exh}} = 4NQ - 1.$

    \item 
    \textbf{ASW-JE~\cite{you2024dft}:} The computation primarily arises from the numerical search over the range space. Using the Fresnel integral function \texttt{fresnel(x)}, available in the JAX library~\cite{jax2018github}, the FLOPs are given as $F_{\text{ASW-JE}} = 6N - 1 + K_a \left( 5N + 694 \times \left( \left\lfloor \frac{\varpi_{\text{max}} - \varpi_{\text{min}}}{\Delta \varpi} \right\rfloor + 1 \right) + 7 \right)$

    \item 
    \textbf{DFT-DNN~\cite{wang2023naivecnn}:} The computational complexity is mainly due to model inference. Following~\eqref{eqn:FLOP}, the FLOPs are
    \begin{equation*}
        \begin{aligned}
            \text{Angle Net: } & \ 26,288N + 6,931,072 + 2047N, \\
            \text{Range Net: } & \ 26,288N + 6,931,072 + 2047Q.
        \end{aligned}
    \end{equation*}
    Therefore, the total FLOPs are $F_{\text{DFT-DNN}} = 52,576N + 13,862,144 + 2047(N + Q) + (4K_b - 1).$

    \item 
    \textbf{DNBT~\cite{qi2023dnbt}:} Similar to DFT-DNN~\cite{wang2023naivecnn}, the computational complexity is derived mainly from model inference. Using~\eqref{eqn:FLOP}, the FLOPs can be expressed as $F_{\text{DNBT}} = 16N^2Q^2 + 6499NQ + \left(4 (\lfloor N/\chi \rfloor)Q + K_b) - 1 \right).$
\end{enumerate}

\begin{figure}[!t]
    \centering
    \includegraphics[width = 1\columnwidth]{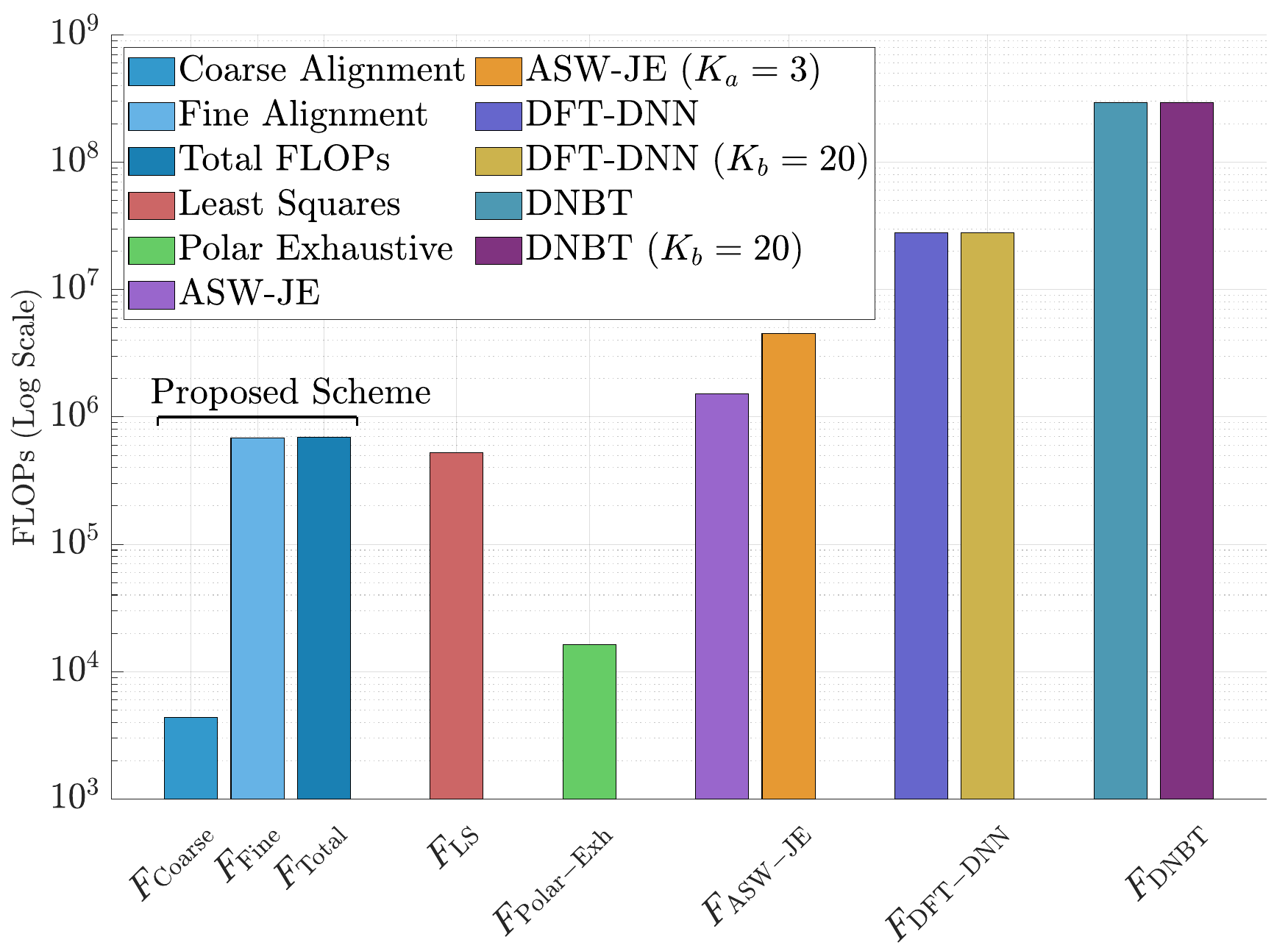}
    \caption{The number of FLOPs in log-scale.}
    \label{fig:flops}
\end{figure}

The computational complexity of each scheme in our simulation setup is illustrated in Fig.~\ref{fig:flops}. The proposed scheme achieves a significantly lower number of FLOPs compared to neural network-based benchmark schemes, such as DFT-DNN~\cite{wang2023naivecnn} and DNBT~\cite{qi2023dnbt}. This reduction is attributed to the effective utilization of domain knowledge during the coarse alignment stage and the use of a reduced-dimensional preprocessed received signal as input, rather than the full received signal. Furthermore, the proposed scheme outperforms ASW-JE~\cite{you2024dft}, which relies on numerical search with Fresnel integrals, in terms of computational efficiency. Remarkably, even without the fine alignment stage, the proposed scheme with only the coarse alignment  delivers superior alignment performance compared to the benchmark schemes, as demonstrated in Section~\ref{subsec:BA_performance}, further validating its effectiveness.  Note that although Polar-Exh~\cite{cui2023rainbow} has the lowest number of FLOPs, as it does not require additional signal processing beyond simple comparisons, its pilot overhead makes it impractical, as shown in Fig.~\ref{fig:achievable_rate}.

Finally, we compare the number of parameters among the neural network-based schemes, DFT-DNN~\cite{wang2023naivecnn} and DNBT~\cite{qi2023dnbt}, in Table~\ref{tab:parameters}. Again, by leveraging domain knowledge, the proposed scheme achieves superior performance even with significantly fewer parameters compared to the baseline schemes. This demonstrates the practicality of the proposed scheme, which utilizes a lightweight model for efficient implementation.

\begin{table}[!ht]
\centering
\caption{Number of parameters of neural networks.}
\label{tab:parameters}
\begin{tabular}{|c|c|}
\hline
\textbf{Model} & \textbf{Number of Parameters} \\ \hline
Proposed Scheme & $81,888$ \\ \hline
Range estimation network~\cite{wang2023naivecnn} & $3,741,392$ \\ \hline
Angle estimation network~\cite{wang2023naivecnn} & $3,987,392$  \\ \hline
DNBT~\cite{qi2023dnbt}  & $134,225,702$ \\ \hline 
\end{tabular}

\end{table}

%\vspace{-20pt}

\section{Conclusion}\label{sec:conclusion}
In this paper, we proposed a novel beam alignment scheme for LoS-dominant near-field communication systems. Our analysis examined the energy spread effect through the correlation between the near-field manifold and the DFT columns, introducing the concept of an  $\epsilon$-approximated signal subspace. Building on this foundation, we developed a hybrid beam alignment scheme that integrates model-based and data-driven approaches. By leveraging the unitary property of the DFT matrix and the defined  $\epsilon$-approximated signal subspace, the coarse alignment stage effectively reduces the search space. Based on this reduced search space, we proposed a DNN-aided fine alignment process to further refine the beam alignment. Simulation results demonstrated that our proposed scheme not only achieves superior alignment performance but also significantly reduces complexity compared to existing methods. Additionally, the proposed scheme ensures backward compatibility with the legacy DFT codebook for far-field signals, allowing for its reuse.

\vspace{-0.2cm}
\bibliographystyle{IEEEtran}  % appearance order
\input{output.bbl}

\end{document}